\documentclass[10pt]{article}
\usepackage{fleqn,ams}
\usepackage[english]{babel}
\usepackage{amssymb,verbatim}
\usepackage{bar,curves,epic,eepic}
\usepackage{verbatim}

\def\stackunder#1#2{\mathrel{\mathop{#2}\limits_{#1}}}

\newcommand{\TwoInt}[1]{%
\stackunder{\hspace{-.2em}\raisebox{-.5ex}{{\small
$#1$}}}{\int\!\!\!\int}}

\newcommand{\Vee}[1]{\stackrel{\vee}{#1}\! }
\begin{document}

\begin{center}

{\bf The statistical dynamics of classic particles ensemble in
gravitational field}\\
      Yu.G.Ignatyev\\
    Kazan State University,\\ Kremleovskaya 18,
Kazan 420008, Russia

\begin{abstract}The article is the translation of authors paper \cite{0},
printed earlier in inaccessible edition and devoted to the
formulation of basic concepts of dynamic description of particles'
statistic ensemble in a gravitational field. Later on, the results
of this article were used by numbers of authors in papers of
relativistic kinetics.\end{abstract}

\end{center}

\section{Introduction}
During last 10 years the independent direction in GRG  - the general
relativistic kinetic theory of matter, bases of which were laid in papers of
N.A.Chernikov \cite{1}- \cite{6} and A.A.Vlasov \cite{7}, segregated and
roughly develops. Methods of general relativistic kinetics are applied in
cosmology \cite{8}- \cite{13}, at researching of gravitational \cite{14}-
\cite{24} and electromagnetic \cite{25}- \cite{31} waves propagation processes
in relativistic plasma, in the theory of Universe gravitational instability
\cite{32}- \cite{37}, as well as in the theory of equilibrium gravitating
plasma configurations \cite{38}- \cite{49}. There were publicized some papers,
devoted to the quantum-field substantiation of classic general relativistic
kinetic theory \cite{50}- \cite{52}, as well as to the quantum-statistical
description of particles' ensemble in the gravitational field \cite{53}-
\cite{55}. The kinetic consideration is necessary in the cases, when high temp
of running processes in the system doesn't allow thermodynamical equilibrium to
establish. However, the strict kinetic model is necessary also for the
substantiation of mac\-ros\-co\-pic hydrodynamics in General Theory of
Relativity. Thus, for instance, the kinetic analysis of cosmological expansion
process display \cite{10},\cite{12}, that processes, leading to the deformation
of equilibrium spectrum of heavy particles, arise in expanding plasma.
According to Weinberg, \cite{56}, this fact can have an important meaning on
the earlier stages of cosmological expansion.

An existing general relativistic theory of gases presents itself
the phenomenological theory (although deeper label, than
hydrodynamics), describing diluted gas, particles of which
interact between themselves on external fields background via
contact binary collisions. The dynamical substantiation of general
relativistic kinetics becomes an urgent problem. In
non-relativistic physics the similar problem was solved in papers
of Bogolyubov, Borne, Grin, Kirkwood, and Evon (BBGKE-chain).
Significant progress in special theory of relativity was reached
by Yu.L.Klimontovich \cite{57}-\cite{59} and R.Balescu
\cite{60}-\cite{62}.

At attempt of constructing of statistical theory in GRG we face
significant difficulties, related with non\-li\-ne\-a\-ri\-ty of
gravitational field. First, non\-li\-ne\-a\-ri\-ty of of Einstein
field equations doesn't allow in strict sense to apply kinetic
theory to the description of interacting particles system
\cite{63}, \cite{64}. In fact, averaging Einstein equations by
different states of particles, we'll obtain
$$\ll G_{ik} \gg = \varkappa\sum_{a}\ll T_{ik}\gg_a,$$
where $\ll \ldots \gg$ means averaging, $a$ - index of particle.
Nonlinear structure of Einstein tensor $G_{ik}$ does not allow us
to write the equality
$$\ll G_{ik}(g) \gg = G_{ik} (\ll g\gg) ,$$
in consequence of that, macroscopic Einstein tensor should be
presented in the form of infinite series
$$\ll G_{ik}(g) \gg = \hskip 3cm$$
\begin{equation}\label{1.1}
=G_{ik} (\ll g\gg) +
\sum_{n=2}^{\infty}\left(\mbox{\parbox{2.25cm}{{\footnotesize
Correlations of local fluctuations of an order n}}}\right).
\end{equation}
Hence, the macroscopic momentum-energy tensor, in strict sense,
doesn't define the macroscopic metric of space-time. This means,
that at constructing of mac\-ro\-s\-co\-pic picture of Universe,
fluctuations of metric tensor are related with correlated
microscopic particles' motions and will stand forward for some
effective momentum-energy tensor. Physically, correlations of
metric tensor local fluctuations can lead to the change of
equation of matter state at high densities (according to Sakharov
\cite{65}, the equation of state can change by radical way). Let's
note, that this moment will play an important role also at
constructing of overall statistic quantum picture of the world.

At constructing of dynamical relativistic theory of interacting
particles we face also with difficulty of re\-tar\-ding field
interactions' description. The finiteness of speed of waves'
propagation lead to the necessity of including to the statistical
system an infinite number of dynamical field variables. In
electrodynamics, mentioned above difficulty, can be overcome, in
principle, due to the linearity of Maxwell field equations. In GRG
the nonlinearity of field equations doesn't allow to present a
metric tensor in form of linear superposition of gravitational
field independent states. Consequently, there is no obvious way of
precise statistical description of interacting particles' ensemble
in GRG. The quantum theory of gravitational field faces with the
same problem.

We suggest an avoiding line of problem solving instead of direct
solution of it. The smallness of gra\-vi\-ta\-ti\-o\-nal
interaction constant allows to hope, that at not very high values
of matter density, when gravitational radiuses of particles don't
overlap, the macroscopic gravitational field in zero approximation
is determined sufficiently well by macroscopic substance and
fields' distribution. In this case it is possible to develop the
perturbation theory by order of correlation parameter smallness,
and to find on that way the corrections to the macroscopic
momentum-energy tensor. At the initial stages of theory
constructing, we will neglect influence of thermal motion of field
on particles' correlations, \footnote{This approximation is
justified by smallness of electromagnetic ($\alpha = e^2/\hbar c =
1/137$) and gravitational ($g_p = Gm^2_p/\hbar c\approx 6\cdot
10^{-39}$) interaction constants, at not very high values of
particles energy.} that corresponds to the exclusion of field
degrees of freedom from the ensemble's distribution function. Such
approximation accounts only retarding particles' interaction and
leads to the kinetic equations without accounting of radiation,
i.e. to the equations of Landau type \cite{66} or to the equations
of Belyaev-Budker \cite{67}. Mentioned programm is particularized
in Ref. \cite{64}.

This article, according to author's opinion, should be considered
not alternatively to completed theory, but rather as a designs of
future statistical theory, as an attempt to collect together the
principles of such theory, and to develop, as far as possible,
it's certain methods. It is hard to predict now, how the right
statistical theory will look like, however, as it seems to Author,
one of the main features of it will be a multitime character of
statistical and field equations. Observed averages at that should
be determined as a synchronized in one or another way of multitime
functions' contraction. This idea presents the key one for given
article.

All denotations, unconditioned in this text specially, are
borrowed from articles \cite{23},\cite{64},\cite{68}.

\section{The Statistical Description Of Medium In Gra\-vi\-tational Field}
In disposal of modern gravitationist there is a sufficient number
of exact and established surely equations of fields and of
particles' motion for basic types of in\-te\-rac\-ti\-ons. Let's
mentally try to write down a self-consistent system of these
equations. The microscopic sources of these fields, determinable
by corresponding ``charges'' and their detail distribution, will
found in the right parts of field equations. In consequence of
pointlike character of particles the field sources are singular.
``Forces'' of interactions, determinable by summary ``tensions''
of fields in points of locations, will found in the right parts of
motions equations. For picture's completeness, it is necessary to
add to the written system of equations, an Einstein equations, in
the right-hand sides of which is situated  summary microscopic
momentum-energy tensor of particles and generated by them fields.
In the latter case, the necessity to write down the equations of
motion, which present themselves differential con\-se\-qu\-en\-ces
of field equations, disappears. In voluminous system of equations,
which appeared before our mental view, all information about
classic particles interaction (including gravitational) is
enclosed. Let's preset on the spacelike hypersurface
\footnote{Here it is necessary to clarify a matter. We'll call
``spacelike hypersurface'' such hypersurface, the normal vector of
which is timelike. Often it is done vise versa: the name, directly
coinciding with the name of type of normal to hypersurface vector,
appropriates to hypersurface. In our opinion it is a violence upon
language. According this scheme, the magnetic field should be
named as a pro-electric only because it's orthogonality to
electric field.} in moment of time $\tau_0$ all ``potentials'' and
their first derivatives by time, and also coordinates and
velocities of all $N$ particles. Then the solutions of this system
of equations will exactly describe an evolution of classical
interacting particles' ensemble.

Let's immerse the observer together with macroscopic apparatus
into ensemble of particles. Let spacelike volume of apparatus is
$\Delta V$. Since observer using this apparatus provides local
measures, and the motion of each particle, falling into apparatus,
is determined by the whole history of this particle's interaction
with others in ranges, un\-con\-t\-rol\-la\-ble by apparatus, then
from observer's point of view the concrete parameters of each
registrable particle (microscopic parameters), and also
macroscopic field parameters are random, unpredictable. However,
if apparatus' volume will be large enough so as to sufficiently
great number of statistically indistinguishable (by given
apparatus) particles will synchronically locate in it, or
sufficiently great number of wave packages will keep within it,
then the observer will fix certain average macroscopic parameters
of particles and fields. For instance, by simple summation of
momentums of statistically indistinguishable particles, lying in
apparatus' volume, the observer can measure the total momentum,
transferring through apparatus by particles of given sort.
Dividing this momentum by apparatus' volume, one will come to the
conception of momentum average density. This value will experience
so small fluctuations (random deviations) in due course, as large
will be the number of particles, synchronically lying in
apparatus. Thus, an observer comes to the idea of measurement of
physical values' certain average macroscopic densities instead of
measurement of these values essentially.

For description of values' densities theorist introduces the
conception of probability density as a function, dependent from
the certain array of microscopic cha\-rac\-te\-ris\-tics of
particles and fields, using which he can calculate any macroscopic
densities of these values. Which requirements should array of
these values satisfy? Let's mentally fulfill the whole space-time
by observers, equipped by macroscopic apparatuses and clocks,
syn\-chro\-ni\-zed by certain way. Mathematically such operation
is equivalent to introduction in $V_4$ of certain timelike vector
field (field of observers), vector lines of which in each point
coincide with the direction of observer's proper time. The
realization of such field of observers with syn\-chro\-ni\-zed
clocks is the semi-geodesic coordinate system \cite{69},
relatively to which observers rest, - synchronic frame of
reference \cite{70}. In time point $\tau_0$ in this frame of
reference we will carry on the measurement of probability density
on the whole hypersurface $V$, orthogonal to the field of
observers, -$\cal{R}\mbox{$_{\tau_0}(V)$}$. Let's require the
r-p\-re\-sen\-ta\-ti\-on $\cal{R}\mbox{$_{\tau_0}(V)$}$ to
determine completely the further evolution of all measurable
values' densities. Probability density for it should be a function
of certain array of values, completely determining the whole
ensemble's dynamics. We'll denote mentioned above values through
dynamic variables, and their full array - through full dynamic
array. Dynamic variables for particles are c-o-di\-na\-tes and
velocities (or momentums) - assignment of these values at defined
fields completely determines particle's trajectory. Dynamic
variables for fields are potentials and their time derivatives.
Here the explanation is necessary. Field equations correspond
themselves equations in partial derivatives in contrast to
particles' equations of motion, which represent themselves
ordinary differential equations. The construction of total dynamic
array of field variables is usually done in such way. Further the
total array of eigenfunctions of vacuum field equations -
$\psi_k(x)$, where $k$ - quantities, composed by certain rule from
quadruple of quantities $k_a$ ($4$-dimensional
Fourier-representation, spherical photon waves etc) should be
determined. In accordance with well-known theorem about
eigenfunctions of Hermit operator, any function can be represented
in form of expansion by $\psi_k(x):\, \sum C_k\psi_k(x)$. General
solution of field equations with source is not an exclusion. Let's
represent time derivatives of potential in form of $\sum
\dot{C_k}\psi_k(x)$, where $\{\dot{C}_k\}$ - recent array of
constants. These infinite arrays $\{ \dot{C_k},C_k\}$
re\-p\-re\-sent themselves exactly the dynamic array of field
variables. The number of field degrees of freedom
co\-n\-se\-qu\-en\-t\-ly is equal to $8\cdot \infty$ (two arrays,
$\{C_k\}$ and $\{\dot{C}_k\}$, each of them includes four arrays
of quantities $k_a$).

Thus, probability density must be a function of type
\begin{equation}\label{2.1}
{\cal D}(\tau ,\tilde{x}_1,\ldots ,\tilde{x}_N,C_1,
\dot{C}_1,\ldots),
\end{equation}
where $\tilde{x}_a$ - totality of dynamic variables of $a-$
particle.

In given article we will neglect ensemble's field degrees of
freedom, i.e. will consider functions (\ref{2.1}), averaged by
field variables. Thereby we exclude free fields (photons) from the
system and limit ourself by accounting of such fields, which are
generated by moving charged particles.

In case of, for instance, electromagnetic interactions such
approximation is fair under the assumption \cite{67}
\begin{equation}\label{2.2}
<{\cal E}>\ll mc^2\sqrt{137L},
\end{equation}
where $L$- Coulomb logarithm. Though the statistic theory with
excluded degrees of freedom is not complete, really relativistic
theory, it nevertheless seems justified and reasonable at first
stages to develop the apparatus of such theory.

In consequence of simple logical construction of statistic theory
with excluded field degrees of freedom, on the basis of this
theory can be testified statistic methods and prepared, perhaps,
more rational and simple formulas for complete statistic theory.

\section{Sources Functions}

On account of exclusive importance of $\delta$-Dirac function in
further calculations we'll pay our fixed attention on it's
features. We will term as an invariant symmetrical double-point
$\delta$-Dirac function, determined on $n$ - di \-men\-si\-o\-nal
Riemannian manifold $R$, a function ${\cal D}(x_1|x_2)$,
possessing next features:
\begin{equation}\label{3.1}
\int\limits_{X_2}{\cal D}(x_1|x_2)F(x_2)dX_2 = \left\{\begin{array}{ll}%
F(x_1); & x_1\in X_2,\\
0; & x_2\not\in X_2,\\
\end{array}\right.
\end{equation}
($X_2\subset R)$; $F(x)$ - random tensor field, $dX =$\\
 $\sqrt{-g(x)}dx^1\ldots dx^n$ - invariant differential of volume
$R$).
\begin{equation}\label{3.2}
{\cal D}(x_1|x_2) = {\cal D}(x_2|x_1);
\end{equation}
\begin{equation}\label{3.3}
{\cal D}(x'_1|x'_2) = {\cal D}(x_1|x_2),
\end{equation}
if $x^{i'} = \varphi^i(x^1,\ldots ,x^n)$- non-degenerate
transformation of coordinates
\begin{equation}\label{3.4}
J = \rm{Det} \left|\!\left|\frac{\partial x^{i'}}{\partial
x^k}\right|\! \right| \not=0.
\end{equation}
Alongside with invariant $\delta$ - function it is possible to
consider also the scalar density $\Delta(x_1|x_2)$, that is what
usually termed $\delta$- Dirac function, -
\begin{equation}\label{3.5}
\Delta(x_1|x_2) = \frac{1}{\sqrt{-g(x_2)}}{\cal D}(x_1|x_2)
\end{equation}
with the law of transformation:
\begin{equation}\label{3.6}
\Delta(x'_1|x'_2) = |J^{-1}(x_2)|\Delta(x_1|x_2).
\end{equation}
Lt us consider an expression of type
\begin{equation}\label{3.7}
\int\limits_RF(x){\cal D}[\psi(x)|0]dx\equiv
\int\limits_RF(x)\Delta[\psi(x)|0]d^nx,
\end{equation}
where $\psi^a(x)$- univalent functions $x$, where
$$J(\psi) = \rm{Det} \left|\!\left|\frac{\partial x^k}{\partial
\psi^a}\right|\! \right| \not=0.$$ Last criterion allows us to
choose $\psi^a$ by way of new coordinates and adduce an integral
to the form
$$\int\limits_{\psi}F[x(\psi)]\Delta(\psi |0)dx|J(\psi)|d^n\psi$$
and, consequently(\ref{3.1}),to obtain formula
\begin{equation}\label{3.8}
\int\limits_RF(x){\cal D}(\psi(x)|0)dx =
\sum_a|J[\psi(x_a)]\,|\,F(x_a),
\end{equation}
that can be written down in form of symbolic rule
\begin{equation}\label{3.8a}
{\cal D}(\psi(x)\,|\,0) = \sum_a\left|J[\psi(x_a)]\right|{\cal
D}(x\,|\, x_a),
\end{equation}
where $x_a$- roots of equation $\psi(x) = 0$;
$$J[\psi(x_a)]= \rm{Det}^{-1}\left|\!\left|\frac{\partial
\psi^6}{\partial x^i_a}\right|\!\right|.$$ Formula (\ref{3.8a}) is
the generalization of well-known property of one-dimensional
$\delta$- Dirac function:
\begin{equation}\label{3.9}
\delta[\varphi(x)] = \sum_a\left|\varphi'(x_a)\right|^{-1}\delta(x
- x_a).
\end{equation}
If in certain frame of reference $R$ is representable in form of
simple product of three-dimensional anisotropic hypersurface $V_k$
and normal to it in each point of coordinate lines' $x_k$
congruence, then $\delta$- Dirac function in this frame of
references can be represented in form of product of invariant on
hypersurface $V_k$ three-dimensional $\delta$- function ${\cal
D}(\tilde{x}_1\,|\,\tilde{x}_2)$ and one-dimensional $\delta$-
function $\delta(\stackunder{k}{x}\!\! _1 |\stackunder{k}{x}\!\!
_2)$
\begin{equation}\label{3.10}
{\cal D}(x_1\,|\,x_2) = {\cal D}( \tilde{x}_1
|\,\tilde{x}_2)\delta(\stackunder{k}{x}\!\! _1
|\stackunder{k}{x}\!\! _2)
\end{equation}
where $\tilde{x}$ - coordinates on the $V_k$. In this frame of
references differential of volume $R$ also represents in form of
product $dX = dV_kdx_k$, where $dV_k =$\\
$\sqrt{-g(x)}d^{n-1}\tilde{x}$ - differential of hypersurface area
$V_k$. Henceforward we will often carry out such an operation in
synchronic frame, when normal vector $k_i$ is timelike. Metric $R$
in this frame has a form
\begin{equation}\label{3.11}
ds^2 = d\tau^2 + g_{\alpha \beta}dx^{\alpha}dx^{\beta}, \quad
(\alpha ,\beta =1,2,3).
\end{equation}
In this case $3$-dimensional hypersurface appears spacelike,
differential of it's area we will denote through $dV$.

Let's consider $\delta$- function derivatives. Extending
integration in (\ref{3.1}) on the whole space and differentiating
this relation, we obtain
$$\frac{\partial F(x_1)}{\partial x^i_1} = \int\limits_RF(x_2)\frac{\partial {\cal
D}(x_1\,|\,x_2)}{\partial x^i_1}dX_2.$$ From the other hand, by
definition
$$\frac{\partial F(x_1)}{\partial x^i_1} = \int\limits_R\frac{\partial F(x_2)}{\partial x^i_2}{\cal
D}(x_1\,|\,x_2)dX_2.$$ Comparing these expressions, we will obtain
the symbolic rule of invariant $\delta$ - Dirac function
differentiation
\begin{equation}\label{3.12}
\frac{\partial}{\partial x^i_1}{\cal D}(x_1\, |\, x_2) = {\cal
D}(x_1\, |\, x_2)\frac{\partial}{\partial x^i_2}.
\end{equation}
Replicating analogous calculations for tensor object $F(x)$, we
will obtain more common differentiation rule
\begin{equation}\label{3.13}
\stackrel{1}{\nabla}_i{\cal D}(x_1\, |\, x_2) = {\cal
D}(x_1\,|\,x_2) \stackrel{2}{\nabla}_i,
\end{equation}
where $\stackrel{a}{\nabla}_i$ - operator of differentiation in
point $x_a.$

Geometric image of particle is timelike world line $x^i = x^i(s_a)
\equiv x^i_a$, along which the certain geometric object
$\omega_a(x_a)\equiv \omega_a(s_a)$, characterizing it's physical
properties, prescribed. We will term this object a source. Only
one tensor object \footnote{Quantum particle in quasi-classic
representation can be conformed also with spinor, determined on
the trajectory.}- velocity vector $u^i_a = dx^i_a/ds_a$ and some
scalar ``charges'' conform to the classic point particle. When
accounting only electromagnetic and gravitational interactions,
charges are equal $- m_a$ (mass), $e_a$ (electric charge). Thus,
classic particle's source can have only a structure of type
$\omega_a(s_a) = \left\{1,e_a,m_a\right\}\times u^{i_1}_a \ldots
u^{in}_a$. Let's determine source's density field
\begin{equation}\label{3.14}
\Omega_a(x) = \int\limits_{\mbox{\parbox{1.8cm}{\footnotesize{
Along whole\\ trajectory}}}}\omega_a(s_a){\cal D}(x\,|\, x_a)ds_a.
\end{equation}
In consequence of invariant $\delta$- function definition,
in\-te\-g\-ra\-ti\-on in (\ref{3.14}) extends tensor properties of
object $\omega_a$ from the particle's trajectory to the whole
manifold $R$, presetting tensor field $\Omega_a(x)$.

Let's introduce into consideration the following sources'
densities, having simple physical meaning:
\begin{equation}\label{3.15}
n^i(x) = \sum n^i_a(x) = \sum_a\int u^i_a(s_a){\cal D}(x\,|\,
x_a)ds_a
\end{equation}
- density vector of particles' number (numerical vector),
$$j^i(x) = \sum e_acn^i_a(x)=$$
\begin{equation}\label{3.16}
 = \sum_ae_ac\int u^i_a(s_a){\cal
D}(x\,|\, x_a)ds_a
\end{equation}
- current's density vector,
$$T^{ik}_p(x) = \sum T^{ik}_a(x) =$$
\begin{equation}\label{3.17}
 = \sum_am_ac^2\int
u^i_a(s_a)u^k_a(s_a){\cal D}(x\,|\,x_a)ds_a
\end{equation}
- tensor of particles' energy-momentum without ac\-c-un\-ting of
interacting fields (tensor of stripped particles'
energy-momentum).

Let's calculate the covariant divergencies from these values.
First we'll consider an expression of type
$$\nabla_in^i_a(x) = \int\limits_Su^i_a(s_a)\frac{\partial {\cal D}(x\,|\,
x_a)}{\partial x^i}ds_a.$$
Let's represent an integral in this expression as a curvi\-linear integral of 2
type, taken along the whole particle's trajectory:

$$\nabla_in^i_a(x) = \int\limits_S\frac{\partial {\cal D}(x\,|\,
x_a)}{\partial x^i}dx^i_a.$$
Also we will account the symbolic rule \ref{3.12}), according to
which operator must act in our case to the unit. Thus, relation
\begin{equation}\label{3.18}
\nabla_in^i_a(x) = \int\limits_Su^i_a(s_a)\frac{\partial {\cal
D}(x\,|\, x_a)}{\partial x^i}ds_a = 0.$$
\end{equation}
always takes place. Mentioned above relation has a form of
conservation law and establishes the evident fact of particle's
existence on it's own trajectory. In consequence of (\ref{3.18})
conservation laws fulfil:
\begin{equation}\label{3.19}
\nabla_in^i(x) = 0;
\end{equation}
\begin{equation}\label{3.20}
\nabla_ij\ \!^i(x) = 0.
\end{equation}

For clarification of physical meaning of these laws we'll obtain
an explicit expression (\ref{3.14}) in synchronic frame of
reference, in which, according to (\ref{3.10}), coordinates
$\tilde{x}$ are given on three-dimensional spacelike $V$. In
con\-se\-qu\-en\-ce of timelikeness of velocity vector $u^i_a$:
$d\tau_a/ds_a \neq 0$, therefore we can proceed from the
integration by proper time $s_a$ in (\ref{3.14}) to integration by
coordinate $\tau_a$. Thus, after integration we will receive
\begin{equation}\label{3.21}
\Omega_a(x) = \omega_a(\tau)\frac{{\cal
D}[\tilde{x}\,|\,\tilde{x}_a(\tau)]}{u^4_a(\tau)},
\end{equation}
where $\tilde{x}_a(\tau)$ means, that coordinates of all particles
are taken in moment $\tau$ of coordinate time. Relation
(\ref{3.21}) can be written also in tensor form. Let's fill up all
$R$ by observer's unitary timelike vector field
\begin{equation}\label{3.22}
(k,k) = 1
\end{equation}
and measure all events by clocks of observers, associated with
field $R$. Space of time, measured by clocks of this observer and
required for infinitely small shift of particle along it's
trajectory, is $d\tau =$ $k_i(x_a)dx^i_a =
k_i(x_a)u^i_a(s_a)ds_a$. Thus, the following relation exists:
\begin{equation}\label{3.23}
ds_a = \frac{d\tau}{(k,u_a)}.
\end{equation}
between integrals $d\tau$ and $ds_a$. At each $\tau =
\mbox{const}$ we will construct the spacelike hypersurface $V$,
orthogonal to the field $k_i$. The equation of this hypersurface
as is well known
$$k_i(x)dx^i = 0.$$
As a result instead (\ref{3.21}) we have
\begin{equation}\label{3.24}
\Omega_a(x) = \omega_a(\tau)\frac{{\cal
D}[\tilde{x}\,|\,\tilde{x}_a(\tau)]}{(k,u_a)_{\tau_a=\tau}}.
\end{equation}
Let's integrate now (\ref{3.19}) all along the hypersurface $V$,
accounting (\ref{3.21}), - as a result we'll obtain
$$\int\limits_V\frac{\partial}{\partial x^i}\left\{\sum
\sqrt{-g(\tilde{x})}\frac{u^i_a(\tau)}{u^4_a(\tau)}{\cal
D}[\tilde{x}\,|\,\tilde{x}_a(\tau)]\right\} d^3\tilde{x} = 0.$$
Let's unfold in details this expression, taking the
dif\-fe\-ren\-tia\-ti\-on by $\tau = x^4 $ outside integration and
introducing velocity $3$- vector of particle on hypersurface
$V$:\\
$v^{\alpha}_a(\tau)/d\tau$
$$\frac{\partial}{\partial \tau}\sum_a\int\limits_V{\cal
D}(\tilde{x}\,|\,\tilde{x}_a(\tau))dV + $$
$$\int\limits_V\frac{\partial}{\partial \tilde{x}^{\alpha}}
\left\{\sqrt{-\tilde{g}}\sum_av^{\alpha}_a(\tau){\cal
D}(\tilde{x}\,|\,\tilde{x}_a(\tau))\right\}d^3\tilde{x}= 0.$$
Let's apply the Gauss formula to the second integral, pro\-cee\-ding to the
integration all along closed two - dimensional surface $\Sigma$, limiting
region $V$:
$$\frac{\partial}{\partial \tau}\sum_a\int\limits_V{\cal
D}(\tilde{x}\,|\,\tilde{x}_a(\tau))dV =$$
\begin{equation}\label{3.25}
= -\TwoInt{\Sigma}\sum_av^{\alpha}_a(\tau){\cal
D}(\tilde{x}\,|\,\tilde{x}_a(\tau))d\Sigma_\alpha .
\end{equation}
In consequence of $\delta$-function definition (\ref{3.1})
integral in the left side of (\ref{3.25}) is equal to the number
of particles, lying in three-dimensional range $V$ in point of
time $\tau$
\begin{equation}\label{3.26}
\sum_a\int\limits_V{\cal D}[\tilde{x}\,|\,\tilde{x}_a(\tau)]dV =
N(\tau).
\end{equation}
Integral in the right side of (\ref{3.25}) is equal to the flux of
number of particles, crossing closed two-dimensional surface
$\sum$, which limits three-dimensional range $V$. Thus,
(\ref{3.25}) posits, that variation of particles' number in range
$V$ is induced only by particles' leaving from that range or
arrival to it - relations (\ref{3.19}), (\ref{3.20}) are
differential forms of particles' and charge's conservation laws
correspondingly.

Let's calculate, finally, the covariant divergence of stripped particles'
energy-momentum tensor. Using re\-la\-ti\-ons (\ref{3.13}), (\ref{3.18}), we'll
obtain
\begin{equation}\label{3.27}
\nabla_kT^{ik}_p(x) = \sum_am_ac^2\int
u^k_a\stackrel{a}{\nabla}_ku^i_a{\cal D}[x\,|\,x_a(s_a)]ds_a.
\end{equation}

\section{Field equations}\label{R4}

Field equations for system of particles, interacting via massive
scalar $(\Phi)$, vector $(A_i)$ and gravitational fields, can be
written down using sources' densities in following form
\begin{equation}\label{4.1}
R^{ik} - \frac{1}{2}Rg^{ik} = \varkappa T^{ik} =\varkappa
(T^{ik}_p + T^{ik}_s + T^{ik}_v),
\end{equation}
\begin{equation}\label{4.2}
F^{ik}_{\,,k} - \mu^2_v A^i = -4\pi \sum e_a\int u^i_a {\cal
D}(x\,|\,x_a)ds_a,
\end{equation}
\begin{equation}\label{4.3}
\stackrel{*}{F} ^{ik}_{\, ,k} = O,
\end{equation}
\begin{equation}\label{4.4}
\Box \Phi + \mu^2_s\Phi = -4\pi \sum q_a\int {\cal
D}(x\,|\,x_a)ds_a,
\end{equation}
where
\begin{equation}\label{4.5}
T^{ik}_p = \sum m_ac^2\left(1 + \frac{q_a\Phi}{m_ac^2}\right)\int
u^i_au^k_a{\cal D}(x\,|\,x_a)ds_a
\end{equation}
- tensor of particles' energy-momentum,
\begin{equation}\label{4.6}
T^{ik}_s = \frac{1}{4\pi}\left\{\Phi^{\,,i}\Phi^{,k} +
\frac{1}{2}g^{ik}(\mu^2_s\Phi^2 - \Phi_{,j}\Phi^{,j})\right\}
\end{equation}
- tensor of massive scalar field's energy-momentum
\footnote{Equation for scalar field does not contain item - $\Phi
R/6$ and therefore at $\mu_s = 0$ is not conformally-invariant.
However even for condormally-invariant field obtained equations of
motion do not differ from (\ref{4.1}), since member $\Phi R/6$ is
compensated by additional members in $T^{ik}_p$.}
$$T^{ik}_v =  \frac{1}{4\pi}\left\{F^{ij}F^{~k}_{j\ .} +\right.$$
\begin{equation}\label{4.7}
\left. + \frac{\mu^2_s}{2}A^iA^k - \frac{1}{4}(F^{jl}F_{jl} +
2\mu^2_vA^jA_j)g^{ik}\right\}
\end{equation}
- tensor of massive vector field's momentums, $e_a$, $q_a$- vector
and scalar charges of particles, $\mu_v = m_vc/\hbar$, $\mu_s =
m_sc/\hbar$; $m_v$, $m_s$ - vector and scalar masses of bosons.

Field equations (\ref{4.1}) - (\ref{4.4}) include ordinary motion
equations. In order to show it, let's apply to both sides of
(\ref{4.1}) the $\nabla_k$ operator and account Bianki identity
law:
$$0 = \sum_{a=1}^{N}\left\{m_ac^2\left(1 +
\frac{q_a\Phi}{m_ac^2}\right)\nabla_k\int u^i_au^k_a{\cal D}
(x\,|\,x_a)ds_a + \right.$$
$$\left. + q_a\Phi_{,k}\int u^i_au^k_a{\cal
D}(x\,|\,x_a)ds_a\right\} +
\frac{1}{4\pi}\left\{F^i_{~.k}(F^{kj}_{~.j} - \mu^2_vA^k) +
\right.$$
\begin{equation}\label{4.8}
\left. + A^iA^k_{~,k}\mu^2_v \right\} +
\frac{1}{4\pi}\left\{\Phi^{,i}(\Box \Phi + \mu^2_s \Phi)\right\}.
\end{equation}
Let's affect via $\nabla_i$ operator to both sides of equation
(\ref{4.2}). In consequence of skew-symmetry of tensor $F^{ik}$ we
will obtain
$$\mu^2_vA^k_{,k} = 4\pi \sum e_a\nabla_k\int u^k_a{\cal D}(x\,|\,x_a)ds_a,$$
and subject to (\ref{3.18}) we'll have the Lorentz calibration at\\
$\mu_v\not\equiv 0.$
\begin{equation}\label{4.9}
A^{k}_{,k} = 0.
\end{equation}
Using this fact, as well as field equations (\ref{4.2}) and
(\ref{4.4}) and relation (\ref{3.27}), we'll adduce (\ref{4.8}) to
the form
$$0 = \sum_{a=1}^{N}\left\{m_ac^2\left(1 +
\frac{q_a\Phi}{m_ac^2}\right)\int
u^k_a\stackrel{a}{\nabla}_ku^i_a{\cal D}(x\,|\,x_a)ds_a -
\right.$$
$$- e_aF^i_{~.k}\int u^k_a{\cal D}(x\,|\,x_a)ds_a - q_a\nabla_k \Phi \left[g^{ik}\int {\cal D}(x\,|\,x_a)ds_a -\right.$$
$$\left. \left. - \int u^i_au^k_a{\cal
D}(x\,|\,x_a)\right]\right\}.$$ Let's choose the proper time of
particle $b$ as a coordinate $x^4$ and integrate the last
expression all along three-dimensional spacelike hypersurface
$V_b$ $(dV_b = \sqrt{-g}\,d\,^3\tilde{x}_b)$. In consequence of
definition of invariant $\delta$- function (\ref{3.1}) we will
obtain motion equations:
$$\frac{{\cal D}p^i_a}{ds_a} =$$
\begin{equation}\label{4.10}
 ={\displaystyle \frac{1}{ 1 + \frac{q_a\Phi}{m_ac^2}}}\left\{\frac{e_a}{c}F^i_{.k}u^k_a +
\frac{q_a}{c}\nabla_k\Phi(g^{ik} - u^i_au^k_a)\right\},
\end{equation}
where $p^i_a = m_acu^i_a$ - particle's momentum. It is
im\-por\-tant to notice, that fields $\Phi$ and $A_i$ are
calculated in point of particle's location, i.e. $A_i = A_i(x_a)$,
$\Phi = \Phi(x_a)$; in the same point also calculated operator
$\stackrel{a}{\nabla}_i$, shift of field coordinates to the point
$x_a$ occurs in consequence of integration of $\delta$- function
all along the spacelike hypersurface.

Let's now discuss the problem with initial conditions for system
(\ref{4.1}) - (\ref{4.4}). We'll incorporate, as we did it in
previous section, the unit timelike field of geodesic observers
$k_i(x)$, and will fix events with the help of their clocks. Let's
construct the spacelike hypersurface $V$, orthogonal to $k_i$. We
will define on this three-dimensional surface in point of time
$\tau_0$ the components of potentials
$\Phi(\tilde{x},\tau_0)\equiv \Phi_0(\tilde{x})$,
$A_i(\tilde{x},\tau_0)\equiv A^0_i(\tilde{x})$; $g_{\alpha
\beta}(\tilde{x},\tau_0)\equiv g^0_{\alpha \beta}(\tilde{x})$ and
their derivatives by time:
$$(\partial_\tau \Phi)_{\tau = \tau_0}\equiv
\dot{\Phi}_0(\tilde{x}), \, (\partial_{\tau}A_i)_{\tau
=\tau_0}\equiv \dot{A}^0_i(\tilde{x}),$$
$$(\partial_{\tau}g_{\alpha \beta})_{\tau = \tau_0}\equiv
\dot{g}^0_{\alpha \beta}(\tilde{x}),$$
as well as the coordinates and momentums of all particles:
$x_a(\tau_0)\equiv x^0_a$, $u_a(\tau_0)\equiv u^0_a $. The
solution of problem with initial conditions for system (\ref{4.1})
- (\ref{4.4}) has a form
\begin{equation}\label{4.11}
\psi(x) = \psi(\tau ,\tilde{x}\,|\,\tilde{x}^0_1,\tilde{u}^0_1,
\ldots ,\tilde{x}^0_N,\Phi_0,\dot{\Phi}_0   \ldots),
\end{equation}
where $\psi(x)$ - certain field functions, depending from values
$\stackrel{0}{g}_{ik}, \tilde{x}^0_a, \Phi_0, \ldots $ as from the
parameters. Thus, field values $(g_{ik}, A_i, \Phi)$ can be
considered as functionals of form
\begin{equation}\label{4.12}
\psi(x) = \psi(\tau ,\tilde{x}\,|\,\tilde{x}_1,\tilde{u}_1, \ldots
,\tilde{x}_N,\tau_0).
\end{equation}
This notation should be understood in the following way: field
values are calculated in point $\tilde{x}$ in instant of time
$\tau_0$, if coordinates and velocities of particles on
hypersurface $V$ in this instant $\tau_0$ were defined by values
$\tilde{x}_a, \tilde{u}_a$.

\section{Many-time formalism}
Field equation's character (\ref{4.1}) - (\ref{4.4}) includes one
feature, besides mentioned in the introduction, which present
problems of invariant construction of statistical model in GRG.
This problem consists in the fact, that motion of each particle is
described by means of invariant proper time $s_a$, while this time
in field equations (\ref{4.1}) must be associated with certain
coordinate $\tau$ in point, where field is calculated. (Let's
recall, that in (\ref{4.12}) values $\tilde{x}_a,\tilde{u}_a$ are
taken in the same point of coordinate time in geodesic frame of
references). This problem, however, can be bypassed, using instead
of field observed values many-time field values
$\stackrel{\vee}{\psi}_N(x\,|\,x_1(s_1), \ldots
,x_N(s_N),u_N(s_N))$, associated with observed $\psi$ with the
help of the rule
$$\psi(x\,|\,\tilde{x}_1,\tilde{u}_1,\ldots
,\tilde{x}_N,\tilde{u}_N) =$$
\begin{equation}\label{5.1}
= S^k_N(s_1,\ldots ,s_N)\Vee{\psi}_N(x\,|\,x_1(s_1),\ldots
,u_N(s_N)),
\end{equation}
where $S^k_N$ - synchronization operator, determined by relation:
\begin{equation}\label{5.2}
S^k_N(s_1,\ldots ,s_N)\Vee{\psi}_N(x\,|\,x_1(s_1),\ldots
,u_N(s_N)) =
\end{equation}
$$
= \int\prod\limits_{a=1}^{N}\delta(s_a -
s_k)ds_ads_k\Vee{\psi}_N(x\,|\,x_1(s_1),\ldots ,u_N(s_N)).$$ Thus,
$$\psi(x,\tau\,|\,x_1(\tilde{\tau}),\ldots ,u_N(\tilde{\tau}))=$$
\begin{equation}\label{5.3}
= \int ds_k\Vee{\psi}_N(x\,|\,x_1(s_k),\ldots ,u_N(s_N)).
\end{equation}
Integration in (\ref{5.2}) - (\ref{5.3}) is carried out along the
whole trajectory of particles and observer. The association
between observer's proper time $s_k$ and coordinate time $\tau$ is
realized by means of relations \footnote{Vector $k$ just chooses
time direction.}
\begin{equation}\label{5.4}
ds_k = k_idx^i.
\end{equation}
From (\ref{5.2}) it follows, that synchronization operator
commutates with any operator, acting solely on coordinates $x$ of
field's ``measuring'' point:
\begin{equation}\label{5.5}
[k_x,S^k_N] = 0.
\end{equation}
Equations for many-time fields we'll write down in form
\begin{equation}\label{5.6}
\Vee{R}_{ik} -
\frac{1}{2}\stackrel{\vee}{R}\stackrel{\vee}{g}_{ik} = {\cal
\kappa}\Vee{T}_{ik},
\end{equation}
\begin{equation}\label{5.7}
\Vee{F}^{ik}_{~,k} - \mu^2_v\Vee{A}^i = -4\pi \sum
e_au^i_a(s_a){\cal D}[x\,|\,x_a(s_a)],
\end{equation}
\begin{equation}\label{5.8}
{F}\raisebox{1.56ex}{\hspace{-1.0em} $\Vee{*}$}\ ^{ik}_{~,k} = 0,
\end{equation}
\begin{equation}\label{5.9}
\Box \Vee{\Phi} + \mu^2_s \Vee{\Phi}\ = -4\pi \sum q_a{\cal
D}[x\,|\,x_a(s_a)],
\end{equation}
where many-time tensor of fields' energy-momentum is constructed
by rules (\ref{4.6}), (\ref{4.7}), but relatively many-time
fields,
$$\Vee{T}^{ik}_p = \hskip 3cm $$
\begin{equation}\label{5.10}%
\sum m_ac^2\Bigl(1 + \frac{q_a\Vee{\Phi}}{m_ac^2}\ \Bigr)
u^i_a(s_a)u^k_a(s_a)\ {\cal D}[x\,|\,x_a(s_a)].
\end{equation}
In consequence of (\ref{5.5}) and (\ref{5.2}) the application of
operator $S^k_N$ to both sides of (\ref{5.6}) - (\ref{5.9}) brings
us to initial field equations (\ref{4.1}) - (\ref{4.4}). From the
definition of many-time field functions is clear, that they should
have $\delta$ - type character.

To make the sense of all told transparent, we'll consider the
example of solution of field equations for massless tensor field
$\psi^{i_1\ldots i_n}(x)$ in flat space-time. Cor\-res\-pon\-ding
many-time equations field equations have form
\begin{equation}\label{5.11}%
\Box \Vee{\psi}^{i_1\ldots i_n}=4\pi \sum q_au^{i_1}_a\ldots
u^{i_n}_a{\cal D}(x\,|\,x_a).
\end{equation}
For solution of problem with initial conditions let's affect on
both sides of (\ref{5.11}) by Fourier operator
\begin{equation}\label{5.12}%
\int\limits_0^{+\infty}dx^4\int\limits_{-\infty}^{+\infty}d^3xe^{i(k,x)},
\end{equation}
as a result we'll obtain the solution
$$\Vee{\psi}_N^{i_1\ldots i_n}(k\,|\,s_1,\ldots ,s_N) = $$
\begin{equation}\label{5.13}%
= - \frac{4\pi}{(k,k)}\sum_+q_au^{i_1}_a \ldots
u^{i_n}_ae^{ik_jx^j_a(s_a)}
\end{equation}
Index ``+'' of symbol $\sum$ means, that only $x^4_a \geq 0$ are
selected (in consequence of (\ref{5.12})). Performing backward
Fourier transformation, after standard calculations we'll find
$$\Vee{\psi}_N^{i_1\ldots i_n}(x\,|\,s_1,\ldots ,s_N) = $$
\begin{equation}\label{5.14}%
= 2\sum_+q_au^{i_1}_a \ldots u^{i_n}_a\delta[(R_a,R_a)],
\end{equation}
where
\begin{equation}\label{5.15}%
R^i_a = x^i_a(s_a) - x^i.
\end{equation}
Let's calculate now the observed field $\psi^{i_1\ldots i_n}(x)$
using relations (\ref{5.13}):
$$\psi^{i_1\ldots i_n}(x\,|\,\tilde{x}_1,\ldots ,\tilde{x}_N) =$$
\begin{equation}\label{5.16}%
= 2\sum_+q_a\int ds_ku^{i_1}_a(s_k) \ldots
u^{i_n}_a(s_k)\delta[(R_a,R_a)],
\end{equation}
where now $x^i_a = x^i_a(s_k)$. For taking of this integral it is
necessary to proceed from integration by variable $s_k$ to
integration by variable $z_a =(R_a,R_a)$. Differentiating, we'll
find the relation
$$\frac{dz_a}{ds_a} = 2(u_a,R_a).$$
Thus, from (\ref{5.16}) we'll receive
$$\psi^{i_1\ldots i_n}(x\,|\,\tilde{x}_1,\ldots ,\tilde{x}_N) =$$
\begin{equation}\label{5.17}%
= \sum_+q_a\frac{u^{i_1}_a \ldots u^{i_n}_a}{(u_a,R_a)}|\,z_a = 0.
\end{equation}
Equation $z_a = 0$ has two roots
\begin{equation}\label{5.18}%
t_a - t = \pm R(t_a)/c,
\end{equation}
where $R(t_a) = |\vec{r}_a(t) - \vec{r}|$. These roots correspond
to retarded and advanced solutions. Since we solve the problem
with initial conditions, then condition $t
> 0$ always fulfills, therefore from (\ref{5.18}) the ``retarded
root'' is selected
\begin{equation}\label{5.18a}%
t_a  +R(t_a)/c = t, \quad t_a \geq 0.
\end{equation}
The relation (\ref{5.18}) is necessary to consider as an equation
for the determination of $t_a(t,\vec{r})$, which after that is
essential to substitute in (\ref{5.17}). At $n = 1 $, $N = 1$
(\ref{5.17}) describes well-known potentials of uniformly moving
charges, at $n = 2$, averaging the distribution of ``charges'',
we'll obtain the solution of linearized Einstein equations with
source \cite{66}.

In order to display how index $k$ of operator $s^k_N$ can be used,
we will consider (\ref{5.16}) in case of $n = 1$, $N = 1$. Then
orientating vector $k$ along $4$ - velocity of particle $u^i_a =
\delta^i_4$, $ds_k = dx^4_a(s_k)$, at once we'll receive from
(\ref{5.16}) the Coulomb's law: $\psi^i = q\delta^i_4/r$.

We have mentioned already, that many-time field functions have
$\delta$-type character. In flat space for massless fields these
functions are equal to zero everywhere, except isotropic cone,
connecting the coordinates of each particle and coordinates of
field's measuring point. In case of massive field, these functions
are different from zero on certain hyperbolic surfaces, lying
inside the light cone, slope ratio of these cones is equal to
relation $v(x,t)/c$, where $v$ - speed of field propagation. It is
obvious by the example of solution of many-time equ\-a\-ti\-ons
for massive field in flat space-time. To which facts the account
of gravitational field leads? Firstly, in gravitational field
isotropic surfaces are not strictly conic; secondly, as it has
been shown in papers of Sibgatullin, Ibrahimov and others, wave
packet in gra\-vi\-ta\-ti\-o\-nal field spreads, that may lead to
the occurrence of statistical tails of initially monochromatic
wave. Strictly speaking, while gra\-vi\-ta\-ti\-o\-nal field
presents, only the high-frequency component of field moves along
isotropic hypersurfaces. As a result of this, low-frequency
com\-po\-nents will arrive to the point of observation later,
summing there with high-frequency com\-po\-nents, ra\-di\-a\-ted
in later instants of time. This will lead to the occurrence of the
continuous spectrum in many-time functions, i.e. to the spreading
of $\delta$-function. The account of gra\-vi\-ta\-ti\-o\-nal
interaction should lead to the same effect. Actually, in light
cones' region of overlap appears gra\-vi\-ta\-ti\-o\-nal
interaction, implication of which will be the appearance in region
of overlap of effective statistical part, which itself will be the
source.

\section{Relativistic Hamilton Dynamics Of Particle}
Most naturally statistic dynamics is formulated in terms of
canonical formalism \footnote{About relativistic canonical
formalism see \cite{13}, \cite{57}, \cite{64}, \cite{71},
\cite{72}.}. Relativistic motion equations of massive particle
have form
\begin{equation}\label{6.1}%
\frac{dx^i}{ds} = \frac{\partial H}{\partial P_i}, \quad
\frac{dP^i}{ds} = - \frac{\partial H}{\partial x_i},
\end{equation}
where $P_i$ - generalized particle's momentum, association of
which with ordinary momentum is
\begin{equation}\label{6.2}%
P^i = mc\frac{dx^i}{ds}\equiv mcu^i,
\end{equation}
is formed by equations (\ref{6.1}), $H(x,P)$ - invariant
Ha\-mil\-ton function. Let $\psi[x(s),P(s)]\equiv \psi(s)$ - is
certain dynamic function, i.e. function of dynamic variables.
Let's cal\-cu\-la\-te this function's derivative along
par\-tic\-le's trajectory
$$\frac{d\psi}{ds} = \frac{\partial \psi}{\partial
x^i}\,\frac{dx^i}{ds} + \frac{\partial \psi}{\partial
P_i}\,\frac{dP_i}{ds}.$$

Thus, subject to (\ref{6.1}) we will obtain
\begin{equation}\label{6.3}%
\frac{d\psi}{ds} = \left\{H,\psi \right\},
\end{equation}
where relativistic Poisson brackets are incorporated
\begin{equation}\label{6.4}%
\left\{H,\psi \right\} = \frac{\partial H}{\partial
P_i}\,\frac{\partial \psi}{\partial x^i} - \frac{\partial
H}{\partial x_i}\,\frac{\partial \psi}{\partial P_i}.
\end{equation}
Poisson brackets possess following algebraic properties:
\begin{equation}\label{6.5}%
 \left\{A,B \right\} = -\left\{B,A\right\};
\end{equation}
$\quad  \left\{\alpha A + \beta B,C\right\} =
\alpha\left\{A,C\right\} + \beta \left\{B,C\right\};$
\begin{equation}\label{6.6}%
 \left\{AB,C \right\} = A\left\{B,C\right\} +
 \left\{A,C\right\}B,
\end{equation}
where $\alpha$, $\beta$ - numbers. Properties (\ref{6.5},
(\ref{6.6})) determine Lie algebra. We can introduce an operator
$[A]$, acting on the random function in such way, that
\begin{equation}\label{6.7}%
[A]B \equiv \left\{A,B\right\}.
\end{equation}
By means of this operator we can built the formal solution of the
equation (\ref{6.3})
\begin{equation}\label{6.8}%
\psi(s) = e^{s[H]}\psi(s_0)\equiv \hat{U}(s)\psi(s_0).
\end{equation}
Since last relations formally do not differ from classical ones
\cite{73}, we may assert, that tran\-s\-for\-ma\-ti\-ons
$\hat{U}(s)$ form Lie group --- the group of canonical
transformations, relatively to which equations (\ref{6.1}) are
invariant. One more important property of Poisson bracket is its
linearity as a differential operator, in consequence of that
\begin{equation}\label{6.9}%
\left\{A, \psi(B)\right\} = \frac{d\psi}{dB}\left\{A,B\right\}.
\end{equation}
In consequence of (\ref{6.5}) and (\ref{6.9}) $[H, \psi(H)] = 0$,
therefore relativistic Hamilton function serves as an integral of
motion equations (\ref{6.1}). We will let this integral equal to
particle's rest momentum
\begin{equation}\label{6.10}%
H(x,P) = mc.
\end{equation}
Hamilton function of charged massive particle, lying in
gravitational, vector and scalar fields, can be represented in
form \cite{13}
$$H(x,P) =$$
\begin{equation}\label{6.11}%
 = \left\{g^{jk}\left(P_j - \frac{e}{c}A_j\right)\left(P_k -
\frac{e}{c}A_k\right)\right\}^{\frac{1}{2}} - \frac{q}{c}\Phi.
\end{equation}
Using (\ref{6.1}), (\ref{6.2}) and (\ref{6.10}) we'll find the
association between $P_j$ and $p^i$:
\begin{equation}\label{6.12}%
 p^i = \frac{\left(P_j - \frac{e}{c}A_j\right)g^{ij}}{{\displaystyle 1 +
 \frac{q\Phi}{mc^2}}}\equiv \frac{P^i - \frac{e}{c}A^i}{{\displaystyle 1 +
 \frac{q\Phi}{mc^2}}}.
\end{equation}
Then in terms of variables $x^i$, $p^i$ Hamilton function
(\ref{6.11}) is equal to
$$H(x,p) = \left\{\left(1 + \frac{q\Phi}{mc^2}\right)\sqrt{(p,p)}
- \frac{q\Phi}{c}\right\},$$ from which subject to (\ref{6.10})
we'll obtain the normalization relation
\begin{equation}\label{6.13}%
(p,p) = m^2c^2
\end{equation}
or $(u,u) = 1$. Using (\ref{6.13}) motion equation (\ref{6.1})
with Hamilton function (\ref{6.11}) during conversion to
non-canonical variables $x^i$, $p^i$ transforms to the form
(\ref{4.10}).

\section{Many-time Hamilton Formalism}
As we have mentioned in section \ref{R4}, field functions in
motion equations (\ref{4.10}) are taken in whereabouts of $a$
particle. In future we will denote the coordinates of $a$-particle
by means of $x^i_a$, and the totality of its dynamic variables
$\left\{x_a,P_a\right\}$ by means of $\xi_a$. Thus, according to
the meaning of equations (\ref{4.1}) - (\ref{4.4}), as well as
(\ref{4.10}) in motion equations of $a$ particle are included
field values (see (\ref{4.12})):
$$\psi(x_a\,|\,\tilde{\xi}_1, \ldots ,\tilde{\xi}_N),$$
where $\xi_a = \xi_a(s_a)$. In accordance to meaning of motion
equations (\ref{4.10}) field values $\psi(x_a)$ include also the
con\-t\-ri\-bu\-ti\-on from proper fields, i.e.
$$\psi(x_a\,|\,\tilde{\xi}_1,\ldots , \tilde{\xi}_a,\ldots
,\tilde{\xi}_N).$$
The account of proper fields, as is well known, leads to
unremovable divergences, therefore we won't cast out proper fields
in motion equations of each particle, supposing that particle
moves as a sampling one in summary field of others.

Then many-time Hamilton function of system of $N$ particles can be
represented in form
\begin{equation}\label{7.1}%
\!\!\!\!H[\xi_1(s_1),\ldots ,\xi_N(s_N)] = \sum
H_a[\xi_a(s_a)\,|\,\tilde{\xi}_1, \ldots ,\tilde{\xi}_N],
\end{equation}
where, for instance, in case of scalar interaction's missing
$$H_a[\xi_a(s_a)\,|\,\tilde{\xi}_1, \ldots ,\tilde{\xi}_N] =$$
$$= \left\{g^{jk}(x_a\,|\,\tilde{\xi}_1, \ldots
,\tilde{\xi}_N)\left[P^a_j -
\frac{e_a}{c}A_j(x_a\,|\,\tilde{\xi}_1, \ldots
,\tilde{\xi}_N)\right]\right.\times$$
\begin{equation}\label{7.2}%
\times \left.\left[P^a_k - \frac{e_a}{c}A_k(x_a\,|\,\tilde{\xi}_1,
\ldots ,\tilde{\xi}_N)\right]\right\}^{\frac{1}{2}}.
\end{equation}
Since Hamilton function is many-time, and these times aren't
connected by any relation, motion equations of particles' ensemble
can be written down in form
$$\frac{dx^i_a}{ds_a} = \frac{\partial H}{\partial P^a_i}
\quad \left(\equiv \frac{\partial H_a}{\partial P^a_i}\right)
\quad (a = 1,\ldots ,N).$$
\begin{equation}\label{7.3}%
\quad \quad \frac{dP^a_i}{ds_a} = -\frac{\partial H}{\partial
x^i_a}\quad \left(\equiv -\frac{\partial H_a}{\partial
x^i_a}\right);
\end{equation}
Field functions' differentiation by variables, placed on the right
from vertical line, is not carried out, since it does not depend
explicitly from the time $s_a$. Exactly such motion equations
during conversion to variables $x^i_a$, $p^i_a$ give equations,
coinciding with (\ref{4.10}).

Let's introduce many-time dynamic function
\begin{equation}\label{7.4}%
\psi_N[\xi_1(s_1), \ldots ,\xi_N(s_N)]\equiv \psi_N(s_1, \ldots
,s_N)
\end{equation}
and timelike field of geodesic observers $k_i(x)$, by clocks of
which $\tau$ we will carry out the synchronization of this
function
\begin{equation}\label{7.5}%
\psi_N(\tau) = S_N(\tau)\psi_N(s_1, \ldots ,s_N),
\end{equation}
where operator of synchronization by particles' ensemble is
determined via relation
$$S_N(\tau)\psi_N(s_1, \ldots ,s_N) =$$
\begin{equation}\label{7.6}%
= \int\prod\limits_{a=1}^{N}\delta[s_a - s^*_a(\tau)]\psi_N(s_1,
\ldots ,s_N)ds_a,
\end{equation}
and $s^*_a(\tau)$ is resulted via solution of motion equations
(\ref{7.3}) relatively to canonical parameter $dx^i_a/ds^*_a =
p^i_a(\tau)$. In synchronic frame of reference $\delta$-functions
in (\ref{7.6}) are transformed to the form $$\delta(x^4_a -
\tau)\frac{d\tau}{ds^*_a}\equiv \delta (x^4_a -
\tau)p^4_a(\tau)/m_ac.$$ Introduced operator differs from operator
$S^k_N$ (\ref{5.2}) in consequence of the fact, that ensemble of
particles is determined by $N$ times, whereas fields of this
ensemble -$N +1$. The connection of these operators is following:
\begin{equation}\label{7.7}%
S^k_N = \int S_N(\tau)d\tau .
\end{equation}
Let's calculate the derivative from $\psi_N(\tau)$. Taking into
account the rule of $\delta$-functions' differentiation, as well
as equations (\ref{7.3}), we'll find
\begin{equation}\label{7.8}%
\frac{d\psi_N}{d\tau}= \int\prod\limits_{a=1}^{N}\delta[s_a -
s^*_a(\tau)]ds_a\sum\limits_{b=1}^{N}\frac{ds^*_b}{d\tau}\left\{H_b,\psi
\right\}_b.
\end{equation}
Let's introduce for convenience the many-time Poisson bracket
\begin{equation}\label{7.9}%
\left\{H,\psi \right\}_{N_s} =
\sum\limits_{a=1}^{N}\frac{ds_a}{d\tau_a}\left(\frac{\partial
H}{\partial P^a_i}\,\frac{\partial \psi}{\partial x^i_a} -
\frac{\partial H}{\partial x^i_a}\,\frac{\partial \psi}{\partial
P^a_i}\right)
\end{equation}
And rewrite (\ref{7.8}) in form
\begin{equation}\label{7.10}%
\frac{d\psi_N}{ds} = S_N(\tau)\left\{H,\psi \right\}_{N_s} .
\end{equation}
The expression (\ref{7.10}) should be understood by following way:
first we calculate the many-time Poisson bracket, next employ the
operator of ensemble synchronization to it .

\section{The Relativistic Phase Space}
At non-degenerate transformations of coordinates $x^i_a$ in
Riemannian space
\begin{equation}\label{8.1}%
x^{i'}_a = \varphi^i(x^1_a,x^2_a,x^3_a,x^4_a)\equiv
\varphi^i(x)|_{x=x_a},
\end{equation}
\begin{equation}\label{8.2}%
J_a = \det \left|\!\left|\frac{\partial \varphi^i}{\partial
x^k}\right|\!\right|_{x=x_a}\neq 0
\end{equation}
components of generalized momentum $P_i$ (as well as $p_i$) are
transformed like covariant vector components
\begin{equation}\label{8.3}%
P^a_{i'} = P^a_k\,\frac{\partial x^k_a}{\partial x^{i'}_a}.
\end{equation}

Overhauling various values of particle's coordinates and in every
point--- various values of momentum, we'll arrive to the
conception of phase space of single particle. This space is fiber,
representing skew product upon base $X_a$ with fiber $P_a(x_a)$
\cite{4}. Base $X_a$ is 4-dimensional Riemannian space, provided
with metric $g_{ik}(x_a)$ and various fields: $\Phi(x_a)$,
$A_i(x_a)$ etc. We will term the base the configuration space of
particle. Fiber $P_a(x_a)$ is the tangent space to $X_A$ with
point of contact $x_a$ and represents the bundle of all kinds of
momentums, applied to the points $x_a$. We will term the fiber
$P_a(x_a)$ a momentum space of particle. The topology of momentum
space is the topology of infinite 4-dimensional
pa\-ral\-le\-le\-pi\-ped, coordinates $P_i$ (or $p_i$ ) of which
possess various values in open interval:
\begin{equation}\label{8.4}%
-\infty < P_i < +\infty.
\end{equation}
It is evident, that (\ref{8.4}) is invariant with respect to
transformations (\ref{8.1}) - (\ref{8.3}). It is necessary to
note, that singling out of conditions only with positive energy on
mass surface (\ref{6.10}) is not always possible (see concerning
to it \cite{72}).

Invariant with respect to transformation of co\-or\-di\-na\-tes
(\ref{8.1}) - (\ref{8.3}) differentials of volumes of
configuration and momentum spaces are \cite{64}
$$dX_a = \sqrt{-g_a}\,dx^1_adx^2_adx^3_adx^4_a,$$
\begin{equation}\label{8.5}%
\qquad  \qquad \quad dP_a =
\frac{1}{\sqrt{-g_a}}dP^a_1dP^a_2dP^a_3dP^a_4.
\end{equation}
In accordance to phase space's definition an invariant
differential of its volume is equal to product
\begin{equation}\label{8.6}%
d\Gamma_a = dx^1_adx^2_adx^3_adx^4_adP^a_1dP^a_2dP^a_3dP^a_4.
\end{equation}
This definition remains also at conversion to momentums $p^a_k$.
If contravariant components $P^i_a$ (or $p^i_a$) are taken for
coordinates of momentum space , then
\begin{equation}\label{8.7}%
dP_a = -\sqrt{-g_a}dP^1_adP^2_adP^3_adP^4_a.
\end{equation}
Many-time phase space of whole particles' ensemble is the direct
product of single particles' phase spaces
\begin{equation}\label{8.8}%
\Gamma = \Gamma_1\otimes \Gamma_2 \otimes \ldots \otimes \Gamma_N.
\end{equation}
According to this definition, the invariant differential of volume
is equal to
\begin{equation}\label{8.9}%
d\Gamma = \prod\limits_{a=1}^{n}d\Gamma_a.
\end{equation}
Thus, phase space of ensemble has the dimensionality $8N$;
ensemble is reflected by point in this space.

\section{Differential And Integral Operations In Phase Space}
Since phase space of ensemble is the direct product of single
particles' phase spaces, it is enough to determine these
operations in phase space of one particle. By the same reason we
will omit particle's index in this section. Let $f(x,p)\equiv
f(\xi)$ - certain function of particle's phase coordinates. Let's
consider an integral of type
\begin{equation}\label{9.1}%
\int\limits_{\Gamma}f(\xi)d\Gamma .
\end{equation}
Since invariant differential of phase space's volume is equal to
product of configuration $(dX)$ and momentum $(dP)$ spaces'
differentials of volumes, (\ref{9.1}) is understood as a skew
integration all along these spaces. It is necessary to emphasize
especially that since $P(x)$ is a fiber, operation
$$\int\limits_{P}dP\int\limits_{X}f(x,P)$$
is not determined and is insensible. Therefore (\ref{9.1}) is
necessary to understood by following way
\begin{equation}\label{9.2}%
\int\limits_{\Gamma}f(\xi)d\Gamma =
\int\limits_{X}dX\int\limits_{P(x)}dPf(x,P),
\end{equation}
where it is forbidden to change the order of integration.

Let $f(\xi)$ is tensor
$$f^{i'_1\ldots i'_n}(x',p') =$$
\begin{equation}\label{9.3}%
= f^{i_1\ldots i_n}(x,p)\,\frac{\partial x^{i'_1}}{\partial
x^{i_1}}\ldots \frac{\partial x^{i'_n}}{\partial x^{i_n}}
\end{equation}
with respect to transformations (\ref{8.1})-(\ref{8.3}). This
tensor can have only structures of type
$$p^{i_1\ldots i_n}f(x,p),\quad
a^{i_1\ldots i_n}(x)f(x,p),$$
or structure of mixed type, where $f(x,p)$ - scalar fun\-c\-ti\-on
in phase space, $a^{i_1\ldots i_n }(x)$ - tensor in configuration.
As it is easy to see, arbitrary scalar with respect to
transformations (\ref{8.1})-(\ref{8.3}) function $f(\xi)$ can be
only the function of variables $a_s(\xi)$:
\begin{equation}\label{9.4}%
f(\xi) = f(a_0,a_1,\ldots ,a_s),
\end{equation}
where $a_s(\xi)$ - $S$-linear forms:
\begin{equation}\label{9.5}%
a_s(\xi) = a_{i_1\ldots i_s}(x)p^{i_1\ldots i_s},
\end{equation}
and $a_{i_1\ldots i_s}(x)$ --- $s$-valent completely symmetrical
tensors in configurational space; in particular $a_0(\xi) =
a_0(x)$- scalar in $X$.

Let's introduce in consideration tensor field in
con\-fi\-gu\-ra\-ti\-o\-nal space
\begin{equation}\label{9.6}%
f^{i_1\ldots i_n}(x) = \int\limits_{P(x)}p^{i_1\ldots
i_n}f(x,p)dp,
\end{equation}
which we will term the $n$ moment about scalar $f(x,p)$.
Integration in (\ref{9.6}) is carried out throughout infinite
4-dimensional parallelepiped.

Let's consider the derivative from scalar $f(x,p)$ by
configurational space
\begin{equation}\label{9.7}%
\frac{\partial f(x,p)}{\partial x^i} \equiv \nabla_if(x,p).
\end{equation}
Let's keep in mind, that $f(x,p)$ has a structure (\ref{9.4}),
then
\begin{equation}\label{9.8}%
\frac{\partial f(x,p)}{\partial x^i} =\sum_{s}\frac{\partial
f}{\partial a_s} \nabla_ia_s(x,p).
\end{equation}
Carrying out the covariant differentiation in (\ref{9.8}) with account of\\
$\partial p_i/\partial x^k = 0$, we'll find
$$\nabla_ia_s(x,p) =  a_{i_1\ldots i_s}p^{i_1\ldots i_s} +$$
\begin{equation}\label{9.9}%
+ a_{i_1\ldots i_s}p^{i_{s+1}}
\Gamma^{(i_1}_{ii_{s+1}}p^{i_2}\ldots p^{i_s)}.
\end{equation}
Thus, values $\nabla_ia_s(x,p)$, and (\ref{9.7}) along with them,
are not covector's components in phase space. This means, that it
is necessary to redefine the operation of covariant
differentiation in phase space. From (\ref{9.9}) it is obvious,
that instead $\nabla_i$ in phase space operator of covariant
differentiation is \cite{7}:
\begin{equation}\label{9.10}%
\widetilde{\nabla}_i = \nabla_i -
\Gamma^j_{ik}p^k\frac{\partial}{\partial p^j}.
\end{equation}
This operator is determined by such way, that
\begin{equation}\label{9.11}%
\widetilde{\nabla}_ip^k = 0.
\end{equation}
We will term an operator $\widetilde{\nabla}_i$  the operator of
covariant differentiation by Cartan \cite{73}, or simply - Cartan
de\-ri\-va\-ti\-ve. Then from (\ref{9.9}) we'll find at once
\begin{equation}\label{9.12}%
\widetilde{\nabla}_ia_s(x,p) = p^{i_1}\ldots p^{i_s}a_{i_1\ldots
i_s,i},
\end{equation}
i.e. according to (\ref{9.8}) we'll obtain the following symbolic
rule of differentiation of functions \cite{67}:
\begin{equation}\label{9.13}%
\widetilde{\nabla}_if(x,p) = \nabla_i[f(x],p),
\end{equation}
which means, that for calculation of Cartan de\-ri\-va\-ti\-ve
from function $f(x,p)$ is enough to calculate the ordinary
covariant derivative from it, temporarily supposing vec\-tor of
momentum at that is covariant constant.

Using the operator of covariant differentiation by Cartan it is
possible to attach to the canonical motion equations (\ref{6.1})
more convenient form \cite{63}:
\begin{equation}\label{9.14}%
\frac{dx^i}{ds} = \frac{\partial H}{\partial P_i},\quad
\frac{{\cal D } P_i}{ds} = -\widetilde{\nabla}_iH.
\end{equation}
In that case Poisson brackets (\ref{6.4}) will be written down in
form
\begin{equation}\label{9.15}%
\frac{d\psi}{ds} = \left\{H,\psi \right\} =
\widetilde{\nabla}_i\psi \frac{\partial H}{\partial P_i} -
\frac{\partial \psi}{\partial P_i}\widetilde{\nabla}_iH.
\end{equation}
In particular, at conversion to ordinary momentums $p^i$ and using
of Hamiltonian's explicit form (\ref{6.11}) we will obtain
\cite{63}, \cite{37} $(\Phi = 0)$:
\begin{equation}\label{9.16}%
\frac{d\psi}{ds} = \left\{p^i\widetilde{\nabla}_i +
\frac{e}{c}F^i_{~.k}p^k\frac{\partial}{\partial p^i} \right\}\psi
.
\end{equation}
In conclusion of this section we will consider an expression of
type
$$\int\limits_{P(x)}\widetilde{\nabla}_i\left\{f(x,p)p^{i_1}\ldots
p^{i_n}\right\}dP.$$ Using rule (\ref{9.13}) and definition
(\ref{9.6}), we will obtain at once an important relation
\cite{67}
\begin{equation}\label{9.17}%
\nabla_if^{i_1\ldots i_n}(x) =
\int\limits_{P(x)}\widetilde{\nabla}_i\left\{f(x,p)p^{i_1}\ldots
p^{i_n}\right\}dP.
\end{equation}

\section{Many-time And Synchronized Distribution Functions}
Let's introduce an invariant many-time distribution function of
$N$ particles' ensemble, ${\cal D}_{N_s}(s_1,\ldots ,s_N)$, which
we will define in such way:
$$dW_{N_s} = \prod\limits_{a=1}^{N}\delta(s_a - s'_a){\cal
D}_{N_s}[\xi_1(s'_1),\ldots ,\xi_N(s'_N)]d\Gamma =$$
\begin{equation}\label{10.1}%
= \left[\mbox{\parbox{2.5cm}{{\footnotesize Probability to find
the ensemle in region $d\Gamma$ with center point
$\left\{\xi_1,\ldots \xi_n\right\}$ }}}\right]
\end{equation}
Why the multi-time probability is introduced by singular way? The
fact is that at any specified array of time $s_1, \ldots ,s_N$ the
probability to find the ensemble on all $N$ $3$-dimensional
spacelike hypersurfaces at whole mo\-men\-tums' array must be
equal to one in consequence of that particles do not disappear
anywhere and do not appear from anywhere. Integration of this one
by $N$ times in infinite limits gives $\infty^N$ degrees and
represents itself an absurd operation. Probability density by its
own meaning is determined on 3-dimensional volume, but not on
4-dimensional one. In accordance to (\ref{10.1}) normalization
relation must fulfil
$$\int\limits_{\Gamma}dW_{N_s} =$$
\begin{equation}\label{10.2}%
\!\!\!\!= \int\limits_{\Gamma}\prod\limits_{a=1}^{N}\delta(s_a -
s'_a){\cal D}_{N_s}[\xi_1(s'_1),\ldots ,\xi_N(s'_N)]d\Gamma = 1,
\end{equation}
where integration is carried out by whole phase space. Function
${\cal D}_{N_s}$ is symmetrical by permutations of iden\-ti\-cal
particles $``a''$ è $``b''$:
$${\cal D}_{N_s}(\xi_1, \ldots ,\xi_a,\ldots ,\xi_b ,\ldots ,\xi_N)
= $$
\begin{equation}\label{10.3}%
= {\cal D}_{N_s}(\xi_1, \ldots ,\xi_b,\ldots ,\xi_a ,\ldots
,\xi_N).
\end{equation}
In consequence of function's ${\cal D}_{N_s}$ many-time character
time coordinates of particles are not connected with anything,
since probability (\ref{10.1}) is calculated in dif\-fe\-rent,
non-correlated by any means times $s_1,\ldots ,s_N$. It is
necessary to synchronize the distribution function so that it has
got its ordinary physical meaning. For syn\-chro\-ni\-za\-ti\-on
we'll introduce in $V_4$, as we did it earlier, a unitary timelike
field of geodesic observers $k_i(x)$, sup\-po\-sing the whole this
field sample, not influencing on the ensemble's dynamics. Then at
conversion to phase spa\-ce's formalism field of observers will
reflect in each particle's phase subspace, like any other field on
$V_4$: $g_{ik}(x),A_i(x),$ $\Phi(x)$ etc; $k_i(x)\to k_i(x_a)$.
For syn\-chro\-ni\-za\-ti\-on it is necessary to set the time
coordinate of particle $x^4_a$ in syn\-chro\-nic frame of
reference equal to the proper (syn\-chro\-nic) time $\tau$ of
observer, situated in the same space point $x^{\alpha}_a$. Then
probability, synchronized by field of observers $k_i(x)$ can be
determined using relation
\begin{equation}\label{10.4}%
dW_\tau = S_N(\tau)dW_{N_s} =
\end{equation}
$$
\!\!\!\!= \prod\limits_{a=1}^{N}\delta[s_a - s^*_a(\tau)]{\cal
D}_{N_s}(s_1, \ldots ,s_N)d\Gamma .$$

Let's introduce also $s$- particle synchronized
dis\-t\-ri\-bu\-ti\-on functions:
$$F_s(\xi_1, \ldots ,\xi_s)\prod\limits_{a=1}^{s}d\Gamma_s =
S_{N-S}(\tau)dW_{N_s} =$$
\begin{equation}\label{10.5}=\prod\limits_{a=1}^{s}\delta(s_a -
s'_a)\int \prod\limits_{b=s+1}^{N}\delta[s_b -
s^*_b(\tau)]\times\end{equation}
$$
\times{\cal D}_{N_s}[\xi_1(s'_1), \ldots
,\xi_s(s'_s),\xi_{s+1}(s'_{s+1})\ldots ]d\Gamma .$$
These functions are many-time by first $s$ particles. Integral
(\ref{10.5}), multiplied on corresponding differential of phase
volume, is equal to probability of first $s$ particles finding in
range $d\Gamma_1 \times \ldots \times d\Gamma_s$ with center in
point $\left\{\xi_1,\ldots ,\xi_s\right\}$ regardless from the
rest $N - s$ particles' position in phase space in point of time
$\tau$. If ensemble consists not from identical particles, it is
necessary to consider besides functions (\ref{10.6}) other arrays.

It is possible in integral (\ref{10.5}) to carry out an
in\-te\-g\-ra\-ti\-on by time variables. For that we'll built
three-dimensional spacelike hypersurfaces $V^a$, orthogonal to
field $k_i(x_a)$ and represent differential of volume of
configurational space $dX_a$ in form
$$dX_a = dV^ad\tau_a.$$
Accounting between $\tau_a$ and $s_a$ (\ref{3.23}) we'll write
down:
\begin{equation}\label{10.6}%
d\tau_a = \frac{(k,p)_a}{m_as}ds_a.
\end{equation}
Then
$$\delta[s_a - s^*_a(\tau)]dX_a\equiv \delta[s_a -
s^*_a(\tau)]dV_ads_a\frac{(k,p)_a}{m_ac}\equiv $$
\begin{equation}\label{10.6a}%
\equiv \delta(\tau_a - \tau)\frac{(k,p)_a}{m_as}dV^ad\tau_a,
\end{equation}
i.e.
\begin{equation}\label{10.6b}%
\delta[s_a - s^*_a(\tau)]ds_a = \delta(\tau_a - \tau)d\tau_a,
\end{equation}
and after integration in (\ref{10.5}) we'll obtain
$$F_s(\xi_1,\ldots ,\xi_s) =
\prod\limits_{a=1}^{s}\delta(s_a-s'_a)\int\prod\limits_{\delta=s+1}^{N}d\Sigma_i\frac{p^idp}{m_ac}\times$$
\begin{equation}\label{10.7}%
\times{\cal D}_{N_s}(\xi'_1,\ldots ,\xi'_s,\xi_{s+1},\ldots
,\xi_N),
\end{equation}
where
\begin{equation}\label{10.8}%
d\Sigma^a_i = dV^ak_i(x_a).
\end{equation}
Most simple meaning is possessed by $1$-s and $2$-s particle
distribution functions, - $F_1(\xi_a)d\Gamma_a$ - is probability
to find $``a''$ particle in differential of volume of phase space.
In consequence of that
\begin{equation}\label{10.9}%
\frac{1}{m_ac}\int\limits_{V^a}d\Sigma^a_i\int\limits_{P_a(x_a)}F_1(\xi_a)p^idP
= 1.
\end{equation}
Let's note also the fact, that normalization relation (\ref{10.2})
for synchronized distribution function has a form
\begin{equation}\label{10.10}%
\int\prod\limits_{a=1}^{N}(m_ac)^{-1} d\Sigma^a_ip^i_adP_a{\cal
D}_{N_s}(\tilde{\xi}_1,\ldots ,\tilde{\xi}_N) = 1.
\end{equation}
For multiplicative function
\begin{equation}\label{10.11}%
{\cal D}_{N_s}(\xi_1,\ldots ,\xi_N) =
\prod\limits_{a=1}^{N}F_1(\xi_a)
\end{equation}
(\ref{10.10}) comes to (\ref{10.9}).

\section{Ensemble-Averages From Dynamic Functions}
Let $\Vee{\psi}[x\,|\,\xi_1(s_a),\ldots ,\xi_N(s_N)]$ - is
many-time field function of coordinates of observation point $x$
and dynamic variables $\xi_1(s_1),\ldots ,\xi_N(s_N)$. According
to meaning of probability (\ref{10.1}), expected value of this
function in condition $\left\{s_1,\ldots ,s_N\right\}$ is equal to
$$\ll\Vee{\psi}(x\,|\,s_1\ldots ,s_N)\gg =$$
\begin{equation}\label{11.1}%
 = \int\limits_{\Gamma}\Vee{\psi}(x\,|\,s_1,\ldots
,s_N)dW_{N_s}(s_1,\ldots ,s_N).
\end{equation}
This record should be understood in following way: in the left
side of (\ref{11.1}) is written down the macroscopic value of
many-time field function $\Vee{\psi}$, measured, when proper time
of first particle is $s_1$, second - $s_2$, -etc. Such values
we'll call many-time ensemble-averages, and operation $\ll \ldots
\gg$-the statistical average. According to the definition,
operator of statistical average commutates with any operator
$K_x$, forcing only on field coordinates $x$:
\begin{equation}\label{11.2}%
[K_x,\ll \ldots \gg] = 0.
\end{equation}
Forcing now on both sides of (\ref{11.1}) operator of
syn\-chro\-ni\-za\-ti\-on by field $k_i$, we'll obtain ensemble -
averages from observed field values
$$\ll\psi(x\,|\, \tilde{\xi}_1,\ldots ,\tilde{\xi}_N)\gg_N =
S^k_N\ll \Vee{\psi}(x\,|\,s_1,\ldots ,s_N)\gg_N=$$
$$\!\!\!\int\limits_{\Gamma}S^k_N\Vee{\psi}[x\,|\,\xi_1(s_1),\ldots
,\xi_N(s_N)]dW_{N_s}[\xi_1(s_1),\ldots ,\xi_N(s_N)]$$
\begin{equation}\label{11.3}%
=\int\limits_{\Gamma}\psi(x\,|\,\tilde{\xi}_1,\ldots
,\tilde{\xi}_N) \times dW_{N_s}[\xi_1(\tau),\ldots
,\xi_N(\tau)].\end{equation}

Hence we see, that operator of statistical average commutates with
operator of synchronization $S^k_N$
\begin{equation}\label{11.4}%
[S^k_N,\ll \ldots \gg_N] = 0.
\end{equation}
Subject to results of previous section (\ref{11.3}) can be written
down in form
$$\ll \psi(x\,|\,\tilde{\xi}_1,\ldots ,\tilde{\xi}_N)\gg _N=$$
\begin{equation}\label{11.5}%
\!\!\!\int\prod\limits_{a=1}^{N}\frac{d\Sigma^a_ip^idP_a}{m_ac}{\cal
D}_{N_s}(\tilde{\xi}_1,\ldots
,\tilde{\xi}_N))\psi(x\,|\,\tilde{\xi}_1,\ldots ,\tilde{\xi}_N).
\end{equation}
In particular, from (\ref{11.5}) and (\ref{10.10}) follows, that
ensemble-average from value, independent from dynamic variables,
is equal to this value itself:
\begin{equation}\label{11.6}%
\ll \psi(x)\gg _N = \psi(x).
\end{equation}

\section{Liouville Equation}
Now we have in our hands all essential apparatus to prosecute the
main problem - an establishment of the equation for the
distribution function. We'll limit ourself with consideration of
only elastic processes of interaction, at which particles do not
disappear and do not born. Let all synchronized observers
completely filling in point of time $s_k\equiv \tau = \tau_0$
activate their clocks. The task of each observer is fixation of
passage time through its point upon particles (if it will appear
hear), and also the establishment of particle's sort. These data
should be recorded by them on cards and after time $\Delta \tau$,
they should finish observations, assemble together and discuss the
information about ensemble's behavior. In language of phase space
observers will reconstruct phase trajectory of ensemble
$\xi_1(\tau),\ldots ,\xi_N(\tau)$. It is obvious, that if
observers had an opportunity to carry out exact measurements, then
they would ascertain by theirs card-file, that in each point of
time $s_k = \tau$ in whole space there were only $N$ particles and
exactly of those sorts, that were in the beginning of the
experiment. Moreover, connecting coordinates of particle's
location points in each point of time, they will make sure in
fact, that particle had moved along continuous line, and had never
disappeared from it. In other words, in each point of time was
realized not only global conservation law of particles, but also
the local one. In language of probability $dW_{\tau}$ this means,
that along the ensemble's phase trajectory it is conserved the
synchronized probability:
\begin{equation}\label{12.1}%
\frac{\delta}{\delta \tau}\left\{dW_{N_s}(\tilde{\xi}_1,\ldots
,\tilde{\xi}_N)\right\} = 0
\end{equation}
and the local law of particles' number conservation and
\begin{equation}\label{12.2}%
\frac{d}{d\tau}\int\limits_{\Gamma}dW_{N_s}(\tilde{\xi}_1,\ldots
,\tilde{\xi}_N) = 0
\end{equation}
- the global law of particles' number conservation. It is obvious,
that (\ref{12.2}) is the consequence of (\ref{12.1}), and
(\ref{12.1}) is the consequence of simple fact of existing of all
particles on their own phase trajectory.

Let's engage now in mathematics. Modifying $dW_{\tau}$ along
trajectory, we'll obtain in consequence of (\ref{10.4})
$$\delta dW_{\tau} = \frac{\delta}{\delta
\tau}\left\{\prod\limits_{a=1}^{N}\delta[s_a - s^*_a(\tau)]{\cal
D}_{N_s}(s_1,\ldots ,s_N)d\Gamma \right\}\delta \tau .$$ Taking
into account, that from $\tau$ depend only arguments of
$\delta$-functions, and remembering simple rule of
$\delta$-functions' differentiation (\ref{3.12}), we'll obtain
$$\delta dW_{\tau} = \prod\limits_{a=1}^{N}\delta[s_a -
s^*_a(\tau)]\delta \tau
\sum_{a=1}^{N}\frac{ds^*_a}{d\tau}\,\frac{\delta}{\delta
s_a}\times$$
$$\times \left\{{\cal D}_{N_s}[\xi_1(s_1),\ldots
,\xi_N(s_N)]d\Gamma \right. =$$
$$=\prod\limits_{a=1}^{N}\delta[s_a -
s^*_a(\tau)]\delta \tau
\left\{\sum_{a=1}^{N}\frac{ds^*_a}{d\tau}\,\frac{\delta}{\delta
s_a}\times \right.$$
$$\left.\times{\cal D}_{N_s}[\xi_1(s_1),\ldots ,\xi_N(s_N)]d\Gamma + \right.$$
\begin{equation}\label{12.3}%
\left. + {\cal D}_{N_s}[\xi_1(s_1),\ldots
,\xi_N(s_N)]\sum_{a=1}^{N}\frac{ds^*_a}{d\tau}\frac{\delta
d\Gamma}{\delta s_a}\right\}.
\end{equation}
In consequence of (\ref{7.3}) we have for $\delta {\cal
D}_{N_s}/\delta s_a$
\begin{equation}\label{12.4}%
\frac{\delta {\cal D}_{N_s}}{\delta s_a} = \left\{H_a,{\cal
D}_{N_s}\right\}_a.
\end{equation}
Let's calculate now the variation $\delta d\Gamma / \delta s_a $.
It is obvious, by virtue of many-time character $d\Gamma$, that
\begin{equation}\label{12.5}%
\frac{\delta d\Gamma}{\delta s_a} = \frac{\delta d\Gamma_a}{\delta
s_b}\Bigl.\prod\limits_{b=1}^{N}\Bigr.' d \Gamma_b,
\end{equation}
where $'$ means, that product is taken by all particles, except
$a$. Variation $\delta d\Gamma_a$ can be calculated in the
following way: let in point of time $s_a$ dynamic variables are
equal to $x^i_a$, $P^a_i$. Carrying out the infinite small shift
along the trajectory, we'll obtain these magnitudes' values in
infinitely near point of time $s'_a = s_a + \delta s_a$:
$$x^{i'}_a = x^i_a + \frac{dx^i_a}{ds_a}\delta s_a + O^2(\delta
s_a),$$
$$P^{'a}_i = P^a_i + \frac{dP^a_i}{ds_a} + O^2(\delta s_a).$$
Using equation (\ref{7.3}) here, we'll obtain
\begin{equation}\label{12.6}%
\qquad \qquad x^{'i}_a = x^i_a + \frac{\partial H_a}{\partial
P^a_i}\delta s_a + O^2(\delta s_a),
\end{equation}
$$P^{'a}_i = P^a_i - \frac{\partial H_a}{\partial x^i_a}\delta s_a +
O^2(\delta s_a).$$ Relations (\ref{12.6}) can be considered as the
transformation of coordinates in differential of volume
$d\Gamma_a$, i.e.
\begin{equation}\label{12.7}%
d\Gamma'_a(s'_a) = J\left(\frac{\partial \xi'_a}{\partial
\xi_a}\right)d\Gamma_a(s_a).
\end{equation}
Carrying out the differentiation of (\ref{12.6}), we'll obtain
elements of determinant
$$\frac{\partial x^{'i}_a}{\partial x^k_a} = \delta^i_k  +
\frac{\delta^2H_a}{\partial x^k_a\partial P^a_i}\delta s_a +
O(\delta s_a),$$
\begin{equation}\label{12.8}%
\qquad \qquad \frac{\partial P^{'a}_i}{\partial P^a_k} =
\delta^k_i - \frac{\delta^2H_a}{\partial P^a_k\partial
x^i_a}\delta s_a + O(\delta s_a).
\end{equation}
As to elements of determinant, which belong to the right upper and
left lower blocks, of type $\partial x^{'i}/\partial P_k$, they
aren't interesting for us, since they make contribution in
Jacobian of superior infinitesimal order, than $o^1(\delta s_a)$.
Let's calculate the Jacobian with the help of (\ref{12.8}), we'll
obtain
\begin{equation}\label{12.9}%
J\left(\frac{\partial \xi'_a}{\partial \xi_a}\right) = 1 +
O(\delta s_a),
\end{equation}
in consequence of that we'll obtain from (\ref{3.15})
\begin{equation}\label{12.10}%
\frac{\delta d\Gamma}{\delta s_a} = 0,
\end{equation}
i.e. ensemble's phase space is conserved during motion along
trajectories of each particle. In classics relations (\ref{12.10})
are named Liouville theorem. Accounting now (\ref{12.1}),
(\ref{12.3}), (\ref{12.4}) and (\ref{12.10}), we'll obtain the
common-relativistic Liouville equation
\begin{equation}\label{12.11}%
\prod\limits_{a=1}^{N}\delta[s_a -
s^*_a(\tau)]\sum_{b=1}^{N}\left\{H_b,{\cal
D}_{N_s}\right\}_b\frac{ds_b}{d\tau_b} = 0.
\end{equation}
Let's note, that Liouville equation allows us to calculate only
synchronized averages by ensemble. However, we can go farther,
following ours many-time ideology, and extend equation
(\ref{12.11}) outside the limits of phase tra\-jec\-to\-ry.
\begin{equation}\label{12.12}%
\sum_{b=1}^{N}\frac{ds_b}{d\tau_b}\left\{H_b,{\cal
D}_{N_s}\right\}_b = 0.
\end{equation}
Such equation as a matter of fact is postulated in papers
\cite{63}, \cite{64}. We'll note, that this equation's many-time
character will not be reflected on values, observed averages by
ensemble, in consequence of synchronization operation's
application.

Let's note, that in consequence of motion equations (\ref{7.3})
along the trajectory of each of particles it's own Hamilton
function is conserved. Therefore on each real trajectory is
fulfilled the normalization relation
\begin{equation}\label{12.13}%
H_a(x,P) = m_ac,
\end{equation}
which describes pairs of non-intersecting hypersurfaces
(pseudospheres) in particles' momentum spaces. Let
$\overline{{\cal D}}_{N_s}(\xi_1,\ldots ,\xi_N)$- solution of
Liouville many-time equation (\ref{12.12}), i.e.
$$\left\{H,\overline{{\cal D}}_{N_s}\right\} = 0.$$
In that case in consequence of Poisson bracket's linearity the
solution of Liouville equation will be also
$${\cal D}_{N_s}(\xi_1, \ldots ,\xi_N)=$$
\begin{equation}\label{12.14}%
 = \overline{{\cal D}}_{N_s}(\xi_1,\ldots ,\xi_N)\psi[H(\xi_1, \ldots ,\xi_N)],
\end{equation}
where $\psi$ - random function. Any real particle with fixed rest
mass can move only all along hypersurface (\ref{12.13}) of
momentums' space, therefore ensemble's distribution function
should have form
$${\cal D}_{N_s}(\xi_1, \ldots ,\xi_N)=$$
\begin{equation}\label{12.15}%
 = \overline{{\cal D}}_{N_s}(\xi_1,\ldots
 ,\xi_N)\prod\limits_{a=1}^{N}\delta(H_a - m_ac),
\end{equation}
where $\overline{{\cal D}}_{N_s}(\xi_1,\ldots \xi_N)$ -
non-singular on pseudospheres (\ref{12.13}) solution of Liouville
equation (\ref{12.12}). It is necessary to emphasize, that
$\delta(H_a - m_ac)$ commutates with Poisson bracket:
\begin{equation}\label{12.16}%
 [\delta(H_a - m_ac),\left\{H_a,{\cal D}_{N_s}\right\}_a] = 0.
\end{equation}
At integration of distribution function there occur in\-teg\-rals
of type
$$\int dP_a\delta(H_a - m_ac)F_a(x_a,P_a),$$
for calculation of which we'll use the property of $\delta$ -
function (\ref{3.9}). An equation (\ref{12.13}) has two roots,
responding to conditions with positive and negative energies
\begin{equation}\label{12.17}%
p_4 = {\cal E}_{\pm}(x,p)/c
\end{equation}
(in synchronic frame of reference roots differ only by sign). Then
$$\delta(H_a - m_ac) = \frac{m_ac}{|p^+_4|}\delta(p^4 -p^4_+) + \frac{m_ac}{|p^-_4|}\delta(p^4
-p^4_-),$$ and we have
$$\int dP_a\delta(H_a - m_ac)F_a(x_a,P_a) =$$
\begin{equation}\label{12.18}%
 = m_ac{\sum}^{\pm}\int
dP^{\pm}_a\overline{F}^{\,\pm}_a(x_a,P_a),
\end{equation}
\begin{equation}\label{12.19}%
dP^{\pm} = \sqrt{-g}\frac{dp^1dp^2dp^3}{|p^{\pm}_4|} =
\frac{1}{\sqrt{-g}}\frac{dp_1dp_2dp_3}{p^4_{\pm}}
\end{equation}
- an invariant differential of volume of corresponding
pseudosphere (see for example \cite{4}), and
$\overline{F}^{\pm}_a(x_a,p_a)$ - functions on the same
pseudosphere, corresponding to conditions of particles with
positive and negative energy. In accordance to (\ref{12.18}) we'll
obtain for them
$$\overline{F}_s(\xi_1,\ldots ,\xi_s) =  \int\prod\limits_{a=s+1}^{N}d\Sigma^a_kp^k(dP^+_a\sigma^+_a +
dP^-_a\sigma^-_a)\times$$
$$\times{\cal D}_{N_s}(\xi_1,\ldots
,\xi_s,\tilde{\xi}_{s+1},\ldots ,\tilde{\xi}_N) =$$
$$=\int\prod\limits_{a=s+1}^{N}d\Sigma^a_k\stackrel{+}{P^k}_adP^+_a\overline{{\cal
D}}_{N_s}(\xi_1,\ldots ,\xi_s,\tilde{\xi}^+_{s+1},\ldots
,\tilde{\xi}^+_N) +$$
$$+\int\prod\limits_{a=s+2}^{N}d
\Sigma^a_j\stackrel{+}{P^j}_adP^+_ad\Sigma^{s+1}_k\stackrel{-}{P^k}_{s+1}dP^-_{s+1}\times$$
\begin{equation}\label{12.20}%
\times\overline{{\cal D}}_{N_s}(\xi_1,\ldots
,\xi_s,\tilde{\xi}^-_{s+1},\tilde{\xi}^+_{s+2},\ldots
,\tilde{\xi}^+_N) +\end{equation}
$$
\int\prod\limits_{a=s+1}^{N}d\Sigma^a_k\stackrel{-}{P^k}_adP^-_a\overline{{\cal
D}}_{N_s}(\xi_1,\ldots ,\xi_s,\tilde{\xi}^-_{s+1},\ldots
,\tilde{\xi}^-_N),
$$
where $\sigma^{\pm}_a$ - operator of sorting by pseudospheres
$$\sigma^{\pm}_a\overline{{\cal D}}_{N_s}(\xi_1,\ldots ,\xi_a,\ldots
,\xi_N)= \overline{{\cal D}}_{N_s}(\xi_1,\ldots
,\xi^{\pm}_a,\ldots ,\xi_N).$$ Sum (\ref{12.20}) contains
$2^{N-s}$ members, obtained by every possible permutations of
pluses and minuses in co\-or\-di\-na\-tes
$\tilde{\xi}_{s+1},\ldots ,\tilde{\xi}_N.$

\section{The Integral Representation of Liouville Equation}
Let us advert to Liouville equation (\ref{12.11}). We will
consider integrals of type
\begin{equation}\label{13.1}%
\int d\Gamma_a\delta(s_a -
s^*_a(\tau))\frac{ds_a}{d\tau}\left\{H_a,{\cal D}_{N_s}\right\}_a.
\end{equation}
We will carry out several transformations with Poisson bracket:
$$
\left\{H_a,{\cal D}_{N_s}\right\}_a = \frac{\partial H_a}{\partial
P^a_i} \frac{\partial {\cal D}_{N_s}}{\partial x^i_a} -
\frac{\partial H_a}{\partial x^i_a} \frac{\partial {\cal
D}_{N_s}}{\partial P^a_i}\equiv$$

$$\equiv {\cal D}_{N_s}\frac{\partial^2H_a}{\partial x^i_a\partial
P^a_i}+ \frac{\partial}{\partial x^i_a}\left({\cal
D}_{N_s}\frac{\partial H_a}{\partial dP^a_i}\right) - {\cal
D}_{N_s}\frac{\partial^2H_a}{\partial P^a_i\partial x^i_a} -$$
$$
-\frac{\partial}{\partial P^a_i}\left({\cal D}_{N_s}\frac{\partial
H_a}{\partial x^i_a}\right)\equiv
$$
\begin{equation}\label{13.2}%
\equiv \frac{\partial }{\partial x^i_a }\left({\cal
D}_{N_s}\frac{\partial H_a}{\partial P^a_i}\right) -
\frac{\partial}{\partial P^a_i}\left({\cal D}_{N_s}\frac{\partial
H_a}{\partial x^i_a}\right).\end{equation}
It easily can be seen, that by means of (\ref{9.15}) Poisson
bracket can be presented in form
$$
\left\{H_a,{\cal D}_{N_s}\right\}_a =
$$
\begin{equation}\label{13.3}%
=\stackrel{a}{\widetilde{\nabla}}_i\left({\cal
D}_{N_s}\frac{\partial H_a}{\partial P^a_i}\right) -
\frac{\partial}{\partial P^a_i}\left({\cal
D}_{N_s}\stackrel{a}{\widetilde{\nabla}}_iH_a\right).
\end{equation}
Let us note, that following relations are always fair
\begin{equation}\label{13.4}%
\qquad \qquad \delta[s_a - s^*_a(\tau)]\frac{ds^*_a}{d\tau} =
\delta(\tau - \tau_a),
\end{equation}
$$\delta[s_a - s^*_a(\tau)]ds_a = \delta(\tau -
\tau_a)d\tau_a.$$ Therefore integral (\ref{13.1}) takes form
$$
J_a =  \int dV^a\int\limits_{P_a}dP_a\times
$$
$$\times\left\{\stackrel{a}{\widetilde{\nabla}}_i\left({\cal
D}_{N_s}\frac{\partial H_a}{\partial P^a_i}\right) -
\frac{\partial}{\partial P^a_i}({\cal
D}_{N_s}\stackrel{a}{\widetilde{\nabla}}_iH_a)\right\}$$ Let us
now take into account the relation (\ref{9.17}), according to
which
\begin{equation}\label{13.5}%
J_a = \int dV^a\stackrel{a}{\nabla}_i\int\limits_{P_a}dP_a{\cal
D}_{N_s}\frac{\partial H_a}{\partial P^a_i}- \int dV^a\times
\end{equation}
$$\times \int\limits_{P_a}dP_a\frac{\partial}{\partial P^a_i}({\cal D}_{N_s}\stackrel{a}{\widetilde{\nabla}}_iH_a)$$
Carrying out $3+1$ -partition in the first integral, we will
obtain
$$\frac{\partial}{\partial \tau}\int dV^a\int\limits_{P_a}dP_a{\cal
D}_{N_s}p^4_{(a)} + \int
dV^a\stackrel{a}{\nabla}_{\alpha}\int\limits_{P_a}dP_a{\cal
D}_{N_s}p^{\alpha}_{(a)}.$$ We will transform the second item in
the last expression using Gauss theorem, and the first item with
account of relation $\partial H_a/\partial p^a_4 = p^4_a =
dx^4/ds_a$ we will adduce to the form:
$$\frac{\partial}{\partial \tau}\int d\Gamma_a\delta(s_a -
s^*_a(\tau)){\cal D}_{N_s}\equiv$$ $$\equiv
\frac{\partial}{\partial \tau}F_{N-1}\equiv
\left(\frac{\partial}{\partial \tau_a}F_{N-1}\right)_{\tau =
\tau_a}$$ -this item turns to zero, since $F_{N-1}$ does not
depend from time coordinate of $a$ particle. We will apply Gauss
theorem also in the second integral in (\ref{13.5}) and carry out
an integration in it all along the mass hypersurface, as a result
we will obtain
$$J_a = \TwoInt{S_a}ds^a_{\alpha}{\sum}^{\pm}\int dP_ap^{\alpha}_a\overline{{\cal D}}_{N_s} -$$
\begin{equation}\label{13.6}%
-\int\limits_{V_a}dV^a_{\alpha}{\sum}^{\pm}\int\limits_{{\cal
P}^{\alpha}}d{\cal P}^k_ap^{\alpha}_a{\cal D
_{N_s}}\stackrel{a}{\widetilde{\nabla }}_kH_a.
\end{equation}
Where external integration in the first member is carried out all
along closed two-dimensional hypersurface $S_a$, limiting the
three-dimensional spacelike hypersurface $\Sigma_a$ of $a$
particle's configurational space; internal integration in the
second member is carried out all along two-dimensional
hypersurfaces ${\cal P}_a$, limiting corresponding pseudospheres
in the momentum space. On these surfaces $$p^2 = -g_{\alpha
\beta}p^{\alpha}p^{\beta}\to +\infty, \quad p_4\to \pm \infty.$$
Generalizing accepted in classical statistics hypothesis \cite{7},
we may suppose, that distribution function of particles' ensemble
${\cal D}_{N_s}$ will sufficiently fast tend to zero at any
particle's configurational coordinates approaching to the
two-dimensional hypersurface $S_a$, limiting spacelike
hypersurface $V_a$ of its configurational space $X_a$, or at
particle's momentums approaching to the two-dimensional
hypersurfaces ${\cal P}_a$, limiting pseudospheres of its momentum
space $P_a$.

At this supposition we will obtain finally
\begin{equation}\label{13.7}%
\int d\Gamma_a \delta[s_a - s^*_a(\tau)]\left\{H_a,{\cal
D}_{N_s}\right\}_a\frac{ds_a}{d\tau} = 0.
\end{equation}
Integrating Liouville equation (\ref{12.11}) sequentially by phase
coordinates of various particles, we will receive the chain of
equations
\begin{equation}\label{13.8}%
\sum_{a=1}^{s}\int\prod\limits_{b=s+1}^{N}d\Gamma^k_b\left\{H_a,{\cal
D} _{N_s}\right\}_a\frac{ds_a}{d\tau} = 0,
\end{equation}
$$(s = 1,\ldots ,N-1),$$
$$d\Gamma^k_b = {\displaystyle \frac{d{\sum}^b_ip^i_bdP_b}{m_bc}} = {\sum}^\pm
d{\Sigma}^b_ip^i_bdP^{\pm}_b\sigma^\pm_b,$$
where
\begin{equation}\label{13.9}%
d\Sigma^b_i = dV^bk_i(x_b),\quad dP^\pm_b = {\displaystyle
\frac{1}{\sqrt{-g_b}}\frac{dp_1dp_2dp_3}{|p^4_\pm|}}.
\end{equation}
Equations (\ref{13.8}) are the integral representation of
Liouville equation (\ref{12.11}). It is necessary to add to them
the equations, obtained by every possible permutations of
identical particles.

\section{Bogolubov's Chain Of Zero Approximation}
In consequence of non-linear character of gravitational
interaction of particles from system (\ref{13.8}) it is impossible
to obtain the chain of equations of Bogolubov's chain type, i.e.
equations, connecting $s$- particle distribution functions with
$s+1$ - particle. It is caused by circumstance, that every
particle's Hamilton function includes coordinates of all the rest
particles. Now let us consider the case, when gravitational
interaction of particles can be neglected in comparison with other
interactions. In this case gravitational field there is the
back\-gro\-und, independent from instant phase coordinates of
particles (i.e. $g_{ik} = g_{ik}(x))$, and Bogolubov's chain can
be obtained for any interactions, satisfying superposition
principle and conserving the integral of rest mass: $H_a = m_ac$.
From methodical considerations we will not go beyond the
consideration of vector interactions.

Accounting the structure of charged particle's Ha\-mil\-to\-ni\-an
in vector and gravitational fields (\ref{7.2}) we will obtain
\begin{equation}\label{14.1}%
\left\{H_a,{\cal D}_{N_s}\right\}_a =
p^i_a\stackrel{a}{\widetilde{\nabla}}_i{\cal D}_{N_s} +
\frac{e_a}{c}F^i _{~ .k}(x_a)p^k_a\frac{\partial {\cal
D}_{N_s}}{\partial p^i_a},
\end{equation}
where $$F_{ik} = F_{ik}(x_a\,|\,\tilde{\xi}_1,\ldots
,\tilde{\xi}_N).$$ In consequence of superposition principle for
vector field
\begin{equation}\label{14.2}%
F_{ik}(x_a\,|\,\tilde{\xi}_1,\ldots ,\tilde{\xi}_N) =
\stackrel{o}{F}_{ik}(x_a) +
\sum_{b=1}^{N}\Bigr.'F_{ik}(x_a\,|\,\tilde{\xi}_b),
\end{equation}
i.e. $\stackrel{o}{F}_{ik}(x_a)$ - external vector field's tensor,
$F_{ik}(x_a\,|\,\tilde{\xi}_b)$ - tensor of vector field, produced
by particle $``a''$ in point $x_b$ and observed in point $x_a$;
mark at sum symbol in (\ref{14.2}) means, that $b\neq a$. Tensor
$\stackrel{\vee}{F}_{ik}(x_a\,|\,\xi_b)$, describing pairwise
interaction can be represented by means of vector potential
\begin{equation}\label{14.3}%
\stackrel{\vee}{F}_{ik}(x_a\,|\,\xi_b) =
\stackrel{a}{\nabla}_i\stackrel{\vee}{A}_k(x_a\,|\,\xi_b) -
\stackrel{a}{\nabla}_k\stackrel{\vee}{A}_i(x_a\,|\,\xi_b),
\end{equation}
which in consequence of (\ref{4.2}) and (\ref{4.3}) satisfies
equations
\begin{equation}\label{14.4}%
\left\{\delta^k_i(\stackrel{a}{\Delta}_2 + \mu^2_v) - R^k_i
(x_a)\right\}\stackrel{\vee}{A}(x_a\,|\,\xi_b) =
\end{equation}
$$= 4\pi
e_bu^i_b{\cal D}(x_a|x_b),$$
\begin{equation}\label{14.5}%
\stackrel{a}{\nabla}_i\stackrel{\vee}{A^i}(x_a|\xi_b) = 0.
\end{equation}
Thus, according to (\ref{14.1}) we have
$$
\left\{H_a,{\cal D }_{N_s}\right\}_a = \left\{p^i_a
\stackrel{a}{\widetilde{\nabla}_i} +
\frac{e_a}{c}\stackrel{o}{F^i}_{\!.k}(x_a)p^k_a\frac{\partial}{\partial
p^i_a }\right\}{\cal D}_{N_s}+$$
\begin{equation}\label{14.6}%
+ \frac{e_a}{c}p^k_a\frac{\partial}{\partial
p^i_a}\sum_{b=1}^{N}F^i_{~.k}(x_a|\tilde{\xi_b}){\cal
D}_{N_s}.\end{equation}
Thus, in consequence of superposition principle the Poisson
bracket (\ref{14.1}) can be represented in form
\begin{equation}\label{14.7}%
\left\{H_a,{\cal D }_{N_s}\right\} =
\left\{\stackrel{o}{H}_a,{\cal D}_{N_s}\right\}_a +
\sum_{b=1}^{N}\left\{H_{ab},{\cal D}_{N_s}\right\}_a,
\end{equation}
where $\stackrel{o}{H}_a$ - operator, dependent only from phase
co\-or\-di\-na\-tes of $a$ particle, $H_{ab}$ - operator,
dependent from couple of particles' ``a'' è ``b'' phase
co\-or\-di\-na\-tes. Such representation is fair not only for
vector interactions, but for any others, satisfying superposition
principle and conserving the integral of rest mass \footnote{For
scalar interactions, for instance,
$$\left\{H_{ab},{\cal D}_{N_s}\right\}_a = -\frac{q_a}{c}\stackrel{a}{\nabla}_i\Phi(x_a|\tilde{\xi}_b)
\frac{\partial {\cal D}_{N_s}}{\partial p^a_i}.$$}. Using
(\ref{14.7}) in (\ref{13.8}), we'll obtain
$$\sum_{a=1}^{s}\frac{ds_a}{d\tau}\int\prod\limits_{b=s+1}^{N}d\Gamma^k_b
\left\{H_a,\overline{{\cal D}}_{N_s}\right\}_a =$$
$$= \sum_{a=1}^s\frac{ds_a}{d\tau}\int\prod_{b=s+1}^{N}d\Gamma^k_b\left\{\stackrel{o}{H_a},
\overline{{\cal D}}_{N_s}\right\}_a +$$
$$
+
\sum_{a=1}^{s}\frac{ds_a}{d\tau}\sum_{b=1}^{s}\Bigr.'\int\prod\limits_{c=s+1}^{N}d\Gamma^k_c
\left\{H_{ab},\overline{{\cal D}}_{N_s}\right\} +$$
$$\sum_{a=1}^{N}\frac{ds_a}{d\tau}\sum_{b=s+1}^{N}\int\prod\limits_{c=s+1}^{N}d\Gamma^k_c\left\{H_{ab},
\overline{{\cal D}}_{N_s}\right\}_a = 0.$$
Using now the definition of $s$-particle distribution
fun\-c\-ti\-ons (\ref{10.5}), we will come to the
common-relativistic Bogolubov's chain \cite{63}
\begin{equation}\label{14.8}%
\sum_{a=1}^{s}\frac{ds_a}{d\tau}\left\{\stackrel{o}{H}_a,\overline{F}_s(\xi_1,\ldots
,\xi_s )\right\}_a +
\end{equation}
$$+
\sum_{a=1}^{s}\frac{ds_a}{d\tau}\sum_{b=1}^{s}\Bigr.'\left\{H_{ab},\overline{F}_s(\xi_1,\ldots
,\xi_s )\right\}_a + \sum_{a=1}^{s}\frac{ds_a}{d\tau}\times$$
$$\sum_{b=s+1}^{N}\int
d\Gamma^k_b\delta[s_b -
s^*_b(\tau)]\left\{H_{ab},\overline{F}\,^b_{s+1}(\xi_1,\ldots
,\xi_s,\xi_b )\right\}_a = 0.$$
Index $b$ at $s+1$- particle distribution function points to the
fact, that not all $F^b_{s+1}$ are equal between themselves in
consequence of particles' nonidentity. In case of iden\-ti\-cal
particles' ensemble we will obtain, carrying out summation in the
last member
\begin{equation}\label{14.9}%
\sum_{a=1}^{s}\frac{dS_a}{d\tau}\left\{\stackrel{o}{H}_a,\overline{F}_s\right\}_a
+
\sum_{a=1}^{s}\frac{dS_a}{d\tau}\sum_{b=1}^{s}\Bigr.'\left\{H_{ab},\overline{F}_s\right\}
+
\end{equation}
$$+ (N -
s)\sum_{a=1}^{s}\frac{dS_a}{d\tau}\int
d\Gamma^k_{s+1}\delta[s_{s+1} - s^*_{s+1}(\tau)]\times$$
$$\times \left\{H_{a,s+1},\overline{F}_{s+1}(\xi_1,\ldots ,\xi_{s+1}
)\right\} = 0.$$ In particular, for vector interaction we have
from (\ref{14.8})
\begin{equation}\label{14.10}%
\sum_{a=1}^{s}\frac{dS_a}{d\tau}\left\{p^i_a\stackrel{a}{\widetilde{\nabla}}_i
+ \frac{e_a}{c}\stackrel{o}{F^i}_{.k}p^k_a\frac{\partial}{\partial
p^i_a}+\right.
\end{equation}
$$\left. + \frac{e_a}{c}p^k_a\frac{\partial}{\partial
p^i_a}\sum_{b=1}^{s}F^i_{~.k}(x_a|\xi_b)\right\}\overline{F}_s
+\sum_{a=1}^{s}\frac{dS_a}{d\tau}\frac{e_a}{c}p^k_a\frac{\partial}{\partial
p^i_a}\times$$
$$\times \sum_{b=s+1}^{N}\int d\Gamma^k_b\delta[s_b -
s^*_b(\tau)]F^i_{~.k}(x_a|\xi_b)\overline{F}_{s+1} = 0,$$
$$(s = 1,2,\ldots ,N-1).$$
These equations jointly with field equation (\ref{4.1}) -
(\ref{4.4}) form the full system of equations, describing gas of
interacting charged particles on the background of
Ri\-e\-man\-ni\-an space.

\section{Conservation Laws In The Statistical Model}
The consequence of Liouville equation (\ref{12.11}) should be the
macroscopic laws of conservation. Let us display this. We will
affect by means of statistical average's operator to the local
macroscopic values $j^i(x)$, $n^i(x)$, $T^{ik}_p$ (\ref{3.15}) -
(\ref{3.17}). At calculation of averages there occur integrals of
type
$$\int d\Gamma_a\delta(s'_a - s_a)\int {\cal
D}(x|x'_a)\psi[\xi_a(s'_a)]ds'_a.$$ Let us find using (\ref{3.24})
values of these integrals:
$$\int \psi(x,p_a)dP_a.$$
Thus, accounting definition (\ref{10.7}), we'll obtain
\begin{equation}\label{15.1}%
\ll n^i(x)\gg_N = \Sigma \frac{1}{m_ac}\int p^i_aF_1 (x,p_a)dP_a,
\end{equation}
\begin{equation}\label{15.2}%
\ll j^i(x)\gg_N = \Sigma \frac{e_a}{m_ac}\int p^i_aF_1
(x,p_a)dP_a,
\end{equation}
\begin{equation}\label{15.3}%
\ll T^{ik}_p(x)\gg_N = \Sigma \frac{1}{m_a}\int p^i_ap^k_a
F_1(x,p_a)dP_a.
\end{equation}
The values of these macroscopic magnitudes coincide with theirs
phenomenological definitions \cite{75}.

For establishment of macroscopic conservation laws we will
consider the first group of integral equations (\ref{13.8}) for
vector interaction
\begin{equation}\label{15.4}%
\int \left\{p^i_a\stackrel{a}{\widetilde{\nabla}}_i\overline{{\cal
D}} _{N_s} + \frac{e_a}{c}F^i_{~.k}p^k_a\frac{\partial
\overline{{\cal D}}_{N_s}}{\partial p^i_a}\right\}\times
\end{equation}
$$\times \frac{ds_a}{d\tau}\prod\limits_{b=1}^{N}d\Gamma^k_b
= 0.$$ Let's integrate these equations all along the momentum
space of $a$ particle, in consequence of (\ref{9.17}) we'll obtain
$$\stackrel{a}{\nabla}_i\int dP_ap^i_a\int
\prod\limits_{b=1}^{N}\Bigr.'d\Gamma^k_b\overline{{\cal D}}_{N_s}
+$$
$$+ \frac{e_a}{c}\int p^k_adP_a\frac{\partial}{\partial p^i_a}\int
\overline{{\cal
D}}_{N_s}F^i_{~.k}\prod\limits_{b=1}^{N}\Bigr.'d\Gamma^k_b = 0.$$
Integrating the second member by parts, accounting the
skew-symmetry of $F^{ik}$ and the definition of many-particle
functions (\ref{10.7}), we will find
\begin{equation}\label{15.5}%
\stackrel{a}{\nabla}_i\int F_1(x_a,p_a)p^idP_a = 0.
\end{equation}
Under the integral (\ref{15.4}) is situated the magnitude, which
is equal, according to (\ref{15.1}) to
\begin{equation}\label{15.6}%
\ll n^i_a(x_a)\gg_N = \int F_1(x_a,p_a)p^i_adP_a(m_ac)^{-1}.
\end{equation}
Substituting coordinates $x_a$ with $x$, we will obtain
\begin{equation}\label{15.7}%
\nabla_i\ll n^i_a(x)\gg_N = 0
\end{equation}
- conservation law of particle. In consequence of (\ref{15.7})
there fulfil the macroscopic laws of conservation of full number
of particles and charges
\begin{equation}\label{15.8}%
\nabla_i\ll n^i(x)\gg_N = 0, \quad \nabla_i\ll j^i(x)\gg_N = 0.
\end{equation}
Let's now multiply (\ref{15.4}) by $p^k_a$ and repeat the whole
made procedure; after that let's sum the result by all particles'
sorts, then we'll obtain
\begin{equation}\label{15.9}%
\stackrel{a}{\nabla}_i\ll T^{ik}_p(x)\gg_N -
\end{equation}
$$- \sum \frac{e_a}{c}\int
p^k_adP_a\int
\prod\limits_{b=1}^{N}\Bigr.'d\Gamma^k_b\overline{{\cal
D}}_{N_s}F^i_{~.k}(x_a|\xi_1,\ldots ,\xi_n) = 0.$$ From
macroscopic equations we have the well-known relation
\begin{equation}\label{15.10}%
\nabla_kT^{ik}_f = -j_kF^{ik}.
\end{equation}
Let's average this relation by ensemble, accounting the fact, that
averaging operator commutates with operator $\nabla_k$:
\begin{equation}\label{15.11}%
\nabla_k\ll T^{ik}_f\gg_N =
\end{equation}
$$=-\sum_{a=1}^{N}\frac{e_a}{c}\int p^k_adP_a\int
\prod\limits_{b=1}^{N}\Bigr.'d\Gamma^k_b\overline{{\cal
D}}_{N_s}F^i_{~.k}.$$ Here we have carried out an integration by
$\delta$- functions of sources, summing (\ref{15.9}) and
(\ref{15.11}), we have the macroscopic law of energy conservation:
$$\nabla_k\ll T^{ik}_p + T^{ik}_f\gg_N = 0.$$

\section{Vlasov Equations}
Let's consider an equation for the one-particle
dis\-t\-ri\-bu\-ti\-on
function $F_1(\xi)$, for that we'll put in (\ref{14.10})\\
$s = 1$. In consequence of definition of operation $\Sigma\
\bigr.'$ the second member in left part of (\ref{14.10}) will
disappear and we will obtain
\begin{equation}\label{16.1}%
\left\{p^i_a\stackrel{a}{\widetilde{\nabla}}_i +
\frac{e_a}{c}\stackrel{o}{F^i}_{.k}p^k_a\frac{\partial}{\partial
p^i_a}\right\}F_1(\xi_a) +
\end{equation}
$$+ \frac{e_a}{c}p^k_a\frac{\partial}{\partial
p^i_a}\sum_{b=2}^{N}\int
d\Gamma^b_k\overline{F}_2(\xi_a,\xi_b)F^i_{~.k}(x_a|\xi_b) = 0.$$
Let us suppose, that distribution function of particles' ensemble
is multiplicative, i.e.
\begin{equation}\label{16.2}%
{\cal D}_{N_s}(\xi_1,\ldots ,\xi_N) = \prod_{a=1}^{N}F_1(\xi_a),
\end{equation}
then
\begin{equation}\label{16.3}%
F_s(\xi_1, \ldots ,\xi_s) = \prod_{a=1}^{N}F_1(\xi_a).
\end{equation}
Correlation between single particles' motion at that is absent,
i.e. interaction - particles interact with each other just via
macroscopic smoothed field $< F_{ik}>_a$. According to the
distribution function's meaning, ave\-ra\-ge value of this field
in point is
\begin{equation}\label{16.4}%
< F_{ik}(x_a)>_a =
\end{equation}
$$=\int\prod\limits_{b=1}^{N}d\Gamma^{k}_bF_{ik}(x_a|\tilde{\xi}_1,\ldots
,\tilde{\xi}_n)\overline{{\cal D}}_{N_s}(\tilde{\xi}_1,\ldots,
\xi_a,\ldots ,\tilde{\xi}_N).$$ Using a superposition principle,
we'll receive from here
\begin{equation}\label{16.5}%
< F_{ik}(x_a)> = \sum_{b=1}^{N}\Bigr.'\int
F_1(\xi_b)F_{ik}(x_a|\widetilde{\xi}_b)d\Gamma^k_b.
\end{equation}
Thus, supposing in (\ref{16.1}) no correlations, we'll obtain
finally
\begin{equation}\label{16.6}%
\left\{p^i\widetilde{\nabla}_i +
\frac{e_a}{c}F^i_{~.k}p^k\frac{\partial}{\partial
p^i}\right\}F_1(x,p) = 0,
\end{equation}
where %
\begin{equation}\label{16.7}%
F_{ik} = \stackrel{o}{F}_{ik}(x_a) + \sum_{b=1}^{N}\Bigr.'\int
F_1(\xi_b)F_{ik}(x_a|\xi_b)d\Gamma^k_b
\end{equation}
- summary vector field. We have obtained so-called collisionless
kinetic equation \cite{64}. Let's act now by average operator
(\ref{11.5}) on both sides of field equations (\ref{4.2}) -
(\ref{4.3}). In consequence of field equations' linearity and
observer's coordinates' independence from particles' phase
coordinates, we will obtain
\begin{equation}\label{16.8}%
-4\pi \Sigma
e_a\int\prod\limits_{b=1}^{N}d\Gamma^k_b\overline{{\cal
D}}_{\tau}\int u^i_a{\cal D}(x|x_a)ds_a=
\end{equation}
$$=\nabla_k\ll F^{ik}(x)\gg_N - \mu^2_{\upsilon}\ll A^i(x)\gg_N,$$
\begin{equation}\label{16.9}%
\nabla_k\ll \stackrel{*}{F}\ \!\!\! ^{ik}(x)\gg_N = 0.
\end{equation}
Carrying out integration by all particles except $a$ one in
(\ref{16.8}), in consequence of one-particle distribution
function's definition we will transform an expression in the right
side of (\ref{16.8}) to the form
$$-4\pi \Sigma e_a\int d\Gamma^k_a\overline{F}_1(\xi_a)\int u^i_a{\cal
D}(x|x_a)dS\equiv $$
$$\equiv -4\pi \Sigma e_a\int\limits_{V_a} d{\sum}^{a}_{k}{\sum}^{\pm}
\int\limits_{P_a}dP_ap^k_aF_1(\xi_a)\int u^i_a{\cal
D}(x|x_a)dS_a.$$ Let's use relation (\ref{3.24}) for invariant
$\delta$ - Dirac function
$$\int u^i_a{\cal D}(x|x_a)dS_a = u^i_a(\tau)\frac{{\cal
D}(\tilde{x}|\tilde{x}_a(\tau))}{(u,k)_a},$$ where $\tau$ - proper
time by clocks of observer $k$, ${\cal
D}(\tilde{x}|\tilde{x}_a(\tau))$ - an invariant $\delta$ -
function on hypersurface $V_a$: then we'll receive
\begin{equation}\label{16.10}%
\ll j^i \gg_N = \sum_{a=1}^{N}\frac{e_a}{m_a}\int
p^i_aF_1(x,p_a)dP_a\equiv
\end{equation}
$$\equiv \sum_{a=1}^{N}e_ac\ll n^i_a\gg_N,$$
where $\ll n^i_a\gg_N$ - vector of particles' number flux density:
\begin{equation}\label{16.11}%
\ll n^i_a(x) \gg_N = \frac{1}{m_ac}\int p^i_aF_1(x,p_a)dP_a
\end{equation}
In full accordance with (\ref{15.6}). Thus, (\ref{16.9}) is
transformed to the form
\begin{equation}\label{16.12}%
\nabla_k\ll F^{ik} \gg_N - \mu^2_{\upsilon}\ll A^i\gg_N =
-\frac{4\pi}{c}\ll j^i\gg_N.
\end{equation}
It is necessary to note, that these are the exact equations,
independent from the supposition about distribution function's
multiplicativity. Representing in form of linear superposition by
fields of single sources
$$\stackrel{\vee}{F}_{ik}(x|s_1,\ldots ,s_N) =
\sum_{a=1}^N\stackrel{\vee}{F}_{ik}(x|s_a)$$ and averaging this
expression
\begin{equation}\label{16.13}%
\ll F_{ik}(x)\gg_N = \sum_{a=1}^N\int
d\Gamma^k_aF_1(\xi_a)F_{ik}(x|\xi_a),
\end{equation}
and then comparing it with (\ref{16.5}), we will obtain
$$\ll F_{ik}(x)\gg_N = < F_{ik}(x)>_a + \int
d\Gamma^k_aF_1(\xi_a)F_{ik}(x|\xi_a).$$ Thus, average macroscopic
field \\ $\ll F_{ik}\gg_N$ differs from average macroscopic $<
F_{ik}>_a$, acting on $a$ particle, only on a value of one
particle's average field. If this difference can be neglected, at
supposition $\ll F_{ik}\gg_N\approx <F_{ik}>_a$, we will come to
the system of common-relativistic Vlasov equations (\ref{16.6}),
(\ref{16.10}), (\ref{16.12}). Let us note, that in consequence of
equation (\ref{16.6}) $4$ - current and flux of particles' number
automatically conserve.
\section{Postscript}
Following stated here program, we can obtain kinetic equations for
one or another type of particles' interactions on the
gravitational field's background. For that it is necessary to
solve the field equations (\ref{4.2}), (\ref{4.3}) (or
(\ref{4.4})) and substitute the solution into Bogolubov's chain,
in which it is essential to account particles' pair correlations
and neglect triple correlations. Such program,
un\-do\-ub\-ted\-ly, will develop in further papers. Not having in
view to analyze here the problem of kinetic equations'
constructing on the gravitational field's background in detail,
we'll point, however, the most striking differences of such theory
from corresponding theory in flat space-time. First of all, in
flat space-time in intervals between collisions particle moves
with constant velocity and does not radiate. In gravitational
field in intervals between collisions particle moves along
geodesic and, as a matter of fact, radiates. This radiation can
lead to the extension of interaction's effective radius. Since
character wa\-ve\-length of bremsstrahlung is $\lambda \sim L$
($L$ - character scale of space curvature), stated effect can play
a noticeable part just under the condition
\begin{equation}\label{17.1}%
L^3n\lesssim 1,
\end{equation}
where $n$ - particles' number density. Secondly,
gra\-vi\-ta\-ti\-o\-nal field redistributes moving particles'
fields (imports anisotropy in these distributions), that also can
influence on final value of dispersion's differential cut set.
And, finally, thirdly, particle's radiation, strictly speaking,
does not spread along geodesic lines (geodesic are just the
trajectories of short-wave quantums $\lambda \ll L$).
Low-frequency radiation ($\lambda \gtrsim L$) can spread with
smaller velocity and, moreover, in a number of cases produce the
static tales, that besides can influence on interaction cut-set's
value. The range, where all these three effects become sufficient,
as it is easy to see, is described by formula (\ref{17.1}). Thus,
gravitation can influence sufficiently upon processes' kinetics of
only sufficiently rarefied medium. The situation, however,
changes, if the medium itself serves the source if
gra\-vi\-ta\-ti\-o\-nal field. In this case $L \sim
1/\sqrt{\varkappa {\cal E}}$ \cite{16} and (\ref{17.1}) has a form
\begin{equation}\label{17.2}
\varkappa {\cal E}\gtrsim n^{2/3}.
\end {equation}

For ultrarelativistic medium with equation of state ${\cal E} =
3nT$ we will receive from here
$$T \gtrsim \frac{1}{3\varkappa n^{1/3}}$$
and, for example, in hot Universe (\ref{17.2}) is realized at
times, lesser than Planck ones: $t\lesssim t_{pe} =
\sqrt{\varkappa \hbar /c}\sim 10^{-43}$ñ. Stated above theory is
in essence the theory on the background of gravitational field.
Let us clarify, whether it is impossible to adapt by some way this
theory for the description of particles' gravitational
interactions. Here we right away run against insoluble obstacles,
which are not limited by obstacles of con\-s\-t\-ruc\-ti\-on of
Einstein equation's general solution (\ref{4.1}). The separation
of such an important for us timelike field of geodesic observers,
or timelike hypersurface, on which initial conditions are preset,
turns into the insensible operation. Actually, for the
se\-pa\-ra\-ti\-on of such field it is necessary to know metric's
microscopic structure in every point of time (which?), but in
order to know this structure, it is necessary to know this
structure on timelike hypersurface (which?). The vicious circle
has become closed. Indeed, we do not need random observer, but
exactly the mac\-ro\-s\-co\-pic one, which moves along geodesics
of macroscopic gravitational field. Con\-se\-qu\-en\-t\-ly, we
need to know this macroscopic field. But we can not define it
again, before we will not determine metric's detail structure,
i.e. while we will not solve completely the problem, in which on
each point of calculation the macroscopic observer appears.

But nevertheless there is an overcome from, see\-ming\-ly,
stalemate, and as a matter of fact, it has been found in papers
\cite{63}, \cite{64} (although it has not been formulated there
sufficiently distinctly). The gravitational field can not be
turned on or off. Exactly this circumstance, which, seemingly,
leads to the vicious circle and to the impossibility to measure
anything until the macroscopic dynamics of ensemble's geometry
will not be known, is the overcome from the deadlock. For any type
of non-gravitational interaction particles' ensemble always has a
``zero condition'', in which macroscopic fields are absent, that
is the consequence of the existence of opposite signs ``charges''.
Therefore any macroscopic field of non-gravitational character can
be turned off. Gravitational ``charges'' of all particles and
fields have the same sign - and it is impossible to turn off the
macroscopic gravitational field. It is impossible to
re\-p\-re\-sent particles' ensemble without gravitational field.
Even in the case of annihilation, concrete particle's
disappearance, its gravitational field does not disappear, since
energy of given particle does not disappear, but solely grades
into another forms of matter. Gravitational field serves more
fundamental, more inert form of matter, than ones or another
particles of ensemble. Thus, the conception of gravitational
field's statistic inertness forms: gravitational field of ensemble
of sufficiently great number of particles has at its heart the
macroscopic character and just small on average macroscopic
con\-s\-ti\-tu\-ent, which is defined by correlated motion of
particles and fields. In mathematical form this conception looks:
\begin{equation}\label{17.3}
g_{ik}(x|\tilde{\xi}_1, \ldots ,\tilde{\xi}_N) = g_{ik}(x) +
\delta g_{ik}(x|\tilde{\xi}_1, \ldots ,\tilde{\xi}_N),
\end {equation}
at that
\begin{equation}\label{17.4}
\ll \delta g_{ik}\gg_N = 0,
\end {equation}
and
\begin{equation}\label{17.5}
\ll \delta g_{ik}\delta g^{jk}\gg_N \ll 1,
\end {equation}
where average is carried out by field of macroscopic observer. It
is seen from this considerations, that (\ref{17.5}) can be broken
only at condition (\ref{17.2}), i.e. at times, smaller than Planck
one, where quantization of gra\-vi\-ta\-ti\-o\-nal field is
already necessary. At condition (\ref{17.5}) the microscopic
constituent of gravitational field can be considered as the small
linear perturbation of metric tensor, i.e. as an ordinary field on
the background and can be described within the limits of proposed
above scheme (see \cite{64}).

What is the location of observer in gravitational fluctuating
world? According to the meaning of medium's statistic description
this observer must be the macroscopic one, i.e. observations
should be carried out in scales lot more than lengths (or times)
of healing of single particles' local gravitational fields. GRG
imposes its indelible impress upon the statistic picture - clocks
and bars on microlabel do not coincide with macroscopic clocks and
bars. Therefore particle's motion in terms of macroscopic
synchronized time will not be geodesic even at non-gravitational
origin forces absence. So, for example, motion of particle with
zero rest mass by clocks and scales of macroscopic observer will
not be described by means of isotropic geodesic line  - dodging in
microscopic gravitational fields particle ``dresses'' by
interaction and in terms of macroscopic observer must looks like
particle with non-zero rest mass. This allows us to suggest
following dependence of graviton's energy from the momentum in
medium:
$${\cal E}^2 = c^2p^2 + m^2_gc^4,$$
where $m_g$ - its effective macroscopic mass. The last relation
greatly resembles the relation between gra\-vi\-ta\-ti\-o\-nal
wave's frequency in medium and its wave vector, determined in
\cite{14}:
$$\omega^2 = k^2c^2 + \omega^2_g.$$
Thus, the statistic picture of graviton's motion in
gra\-vi\-ta\-ting medium can give $m_g \sim \omega_g$ and,
thereby, determine physical correspondence between these two
pictures.

In conclusion author passes his appreciation to
\\ \fbox{G.G.Ivanov}, numerous and interesting discussions with whom
in the course of more than five years cooperated extensively to
the gradual crystallization of many con\-cep\-ti\-ons, developed
in this paper. Especially this relates to the conception of
observer.

\end{document}